\documentclass[twocolumn,aps,prd,showpacs,nofootinbib,floatfix,eqsecnum]{revtex4}
\usepackage{amsmath,latexsym,graphicx,multirow,longtable} 
\begin{document}
\title{Towards adiabatic waveforms for inspiral into Kerr black
holes:\\ I. A new model of the source for the time-domain perturbation
equation}
\author{Pranesh A.\ Sundararajan${}^{1}$}
\author{Gaurav Khanna${}^{2}$}
\author{Scott A.\ Hughes${}^{1}$}
\affiliation{${}^{1}$Department of Physics and MIT Kavli Institute,
MIT, 77 Massachusetts Ave., Cambridge, MA 02139 \\ ${}^{2}$Department
of Physics, University of Massachusetts, Dartmouth, MA 02747}
\date{\today}
\begin{abstract}
We revisit the problem of the emission of gravitational waves from a
test mass orbiting and thus perturbing a Kerr black hole. The source
term of the Teukolsky perturbation equation contains a Dirac delta
function which represents a point particle. We present a technique to
effectively model the delta function and its derivatives using as few
as four points on a numerical grid. The source term is then
incorporated into a code that evolves the Teukolsky equation in the
time domain as a (2+1) dimensional PDE. The waveforms and energy
fluxes are extracted far from the black hole. Our comparisons with
earlier work show an order of magnitude gain in performance (speed)
and numerical errors less than $1\%$ for a large fraction of parameter
space. As a first application of this code, we analyze the effect of
finite extraction radius on the energy fluxes. This paper is the first
in a series whose goal is to develop adiabatic waveforms describing
the inspiral of a small compact body into a massive Kerr black hole.
\end{abstract}
\pacs{04.25.Nx, 04.30.Db, 04.30.-w}
\maketitle

\section{Introduction}
\label{sec:intro}

\subsection{Background}

The extreme mass ratio limit of binary systems --- binaries with one
mass far smaller than the other --- has been a special focus of
research in gravitation in recent years.  This is in part because this
problem is, at least formally, particularly clean and beautiful: the
mass ratio allows us to treat the binary as an exact black hole
solution plus a perturbation due to the secondary mass.  Perturbative
techniques can be used to analyze the system, making it (in principle
at least) much more tractable than the general two-body problem in
general relativity.

This limit is also of great astrophysical interest, as it perfectly
describes {\it capture binaries}: binary systems created by the
capture of stellar mass compact objects onto relativistic orbits of
massive black holes in galaxy cores.  Post formation, the evolution of
such binaries is driven by gravitational-wave (GW) emission --- the GW
backreaction circularizes and shrinks the binaries, eventually driving
the smaller body to plunge and merge with its larger companion.  Such
events are now believed to be relatively abundant (see Ref.\
{\cite{hopman}} for up-to-date discussion and review of the relevant
literature).  Since the last year or so of the inspiral is likely to
generate GWs that lie in the low-frequency band of space-based GW
antennae such as {\it LISA} {\cite{lisa}}, {\it extreme mass ratio
inspirals} (or EMRIs) are key targets for future GW observations.

This paper is the first in a series whose aim is to develop {\it
adiabatic EMRI waveforms}.  ``Adiabatic'' refers to the fact that they
are computed using an approximation to the true equations of motion
that takes advantage of the nearly periodic nature of the smaller
body's motion on ``short'' timescales.  This approximation fails to
capture certain important aspects of the binary's evolution.  In
particular, adiabatic waveforms only incorporate {\it dissipative}
effects of the small body's perturbation --- effects which cause
radiation of energy and angular momentum to distant observers and down
the hole, driving the orbit to decay.  {\it Conservative} effects ---
effects which conserve energy and angular momentum, but push the orbit
away from the geodesic trajectory of the background spacetime --- are
missed in this approach.  It has been convincingly demonstrated
{\cite{ppn05}} that conservative effects change orbital phasing in a
way that could be observationally significant.  The dissipative-only
adiabatic approach to EMRI waveform generation is thus, by
construction, somewhat deficient.

In our view, this deficiency is outweighed by the fact that it will
produce waveforms that capture the spectral features of true waveforms
--- a complicated shape ``colored'' by the three fundamental orbital
frequencies and their harmonics.  Also, the adiabatic approach is
likely to produce these waveforms on a relatively short timescale.
Though not perfectly accurate, adiabatic waveforms will be an
invaluable tool in the short term for workers developing a data
analysis architecture for measuring EMRI events.  In the long term,
these waveforms may even be accurate enough to serve as ``detection
templates'' for EMRI events.  {\it Measuring} the characteristics of
EMRI sources will require matching data with as accurate a model as
can be made, and over as long a timespan as possible --- perhaps a
year or more.  By contrast, {\it detecting} EMRI events does not
require matching a signal with a template for such a long time
{\cite{gair04}}.  For the short integration times needed for
detection, work in progress indicates that conservative effects do not
shift the phase so badly that the signal fails to match a template.
What shift does accumulate due to conservative effects can be
accomodated by systematic errors in source parameters, allowing
detection to occur.  (This is discussed in Appendix A of Ref.\
{\cite{dfh05}}, Chapter 4 of Ref.\ {\cite{favata_thesis}}, and Refs.\
{\cite{favata_inprep,fh_inprep}}, currently in preparation.)

\subsection{Our approach to adiabatic inspiral}

The approach which we advocate for building adiabatic waveforms uses a
hybrid of frequency-domain and time-domain perturbation theory
techniques.  These two techniques have complementary strengths and
weaknesses; by combining the best features of both toolsets, we hope
to make waveforms that are as accurate as possible.  Though a
diversion from the main topic of this paper, this approach is a key
motivation for our work.  We thus ask the reader to indulge us as we
briefly describe our rationale.

In the adiabatic limit and neglecting conservative effects, the
separation of timescales means that orbits are, to high accuracy,
simply geodesic trajectories of the spacetime on short timescales.
The orbital decay that is driven by backreaction amounts to the system
evolving from one geodesic orbit to another.  Computing the effect of
radiation reaction thus amounts to computing the sequence of orbits
through which the system passes en route to the final plunge of the
smaller body into the large black hole {\cite{hdff05}}.

A geodesic orbit is characterized (up to initial conditions) by three
constants: energy $E$; axial angular momentum $L_z$; and ``Carter
constant'' $Q$ (see, e.g., {\cite{mtw}}, Chap.\ 33).  Computing this
sequence of orbits is equivalent to computing the rate at which these
constants change due to radiative backreaction.  In this picture, it
is useful to regard each orbit $(E,L_z,Q)$ as a point in an orbital
phase space, and to regard the rates at which they evolve, $(\dot E,
\dot L_z, \dot Q)$, as defining a tangent vector to the trajectory an
evolving system traces through this phase space.  Adiabatic radiation
reaction thus amounts to calculating this tangent at all orbits.

In the extreme mass ratio limit, the smaller body moves very slowly
through orbit space --- it spends many orbits in the vicinity of each
$(E,L_z,Q)$.  This slow evolution means that the tangent vector is
most accurately represented by the {\it average} rate at which these
constants evolve: $(\langle\dot E\rangle, \langle\dot L_z\rangle,
\langle\dot Q\rangle)$, where the angle brackets denote an appropriate
averaging with respect to the orbits.  Such an averaging is defined in
Ref.\ {\cite{dfh05}}.

Once adiabatic radiation reaction data has been found for all orbits,
it is straightforward to choose initial conditions and compute the
worldline ${\bf z}(t)$ which an inspiralling body follows.  In this
framework, it is just a geodesic worldline with the constants slowly
evolving:
\begin{equation}
{\bf z}(t) = {\bf z}_{\rm geod}[E(t), L_z(t), Q(t)]\;.
\end{equation}
This worldline can then be used to build the source term for the wave
equation, allowing us to compute the gravitational waves generated as
the small body spirals in.  We note here that this approach is
conceptually identical to the ``kludge'' presented in Ref.\
{\cite{kludge}}.  Indeed, the almost unreasonable success of kludge
waveforms served as an inspiration for this formulation of
inspiral\footnote{The major difference between the hybrid inspiral
described here and the kludge is that the hybrid inspiral aims to
correctly solve a wave equation at all points along the orbit.  The
kludge instead uses a physically motivated approximate wave formula
based on variation of the source's multipole moments, defined in a
particular coordinate system.}.

In the hybrid approach, a frequency-domain code would be used for the
adiabatic radiation reaction, and a time-domain code used to generate
the waves from a small body following the worldline that radiation
reaction defines.  Since any function built from bound Kerr black hole
orbits has a spectrum that is fully described by three easily computed
frequencies and their harmonics {\cite{schmidt02, dh04}}, the
averaging needed in this prescription is extremely fast and easy to
compute in the frequency domain.  Many harmonics may be needed, but
each harmonic is independent of all others.  Frequency-domain codes
are thus easily parallelized and the calculation can be done very
rapidly.  In the time domain, averaging is much more cumbersome --- a
geodesic orbit and the radiation it generates must be followed over
many orbits to insure that all beatings between different harmonics
have been sampled.  Convergence to the true average for quantities
like $\langle \dot E\rangle$ will be slow for generic (inclined and
eccentric) Kerr black hole orbits (the most interesting case,
astrophysically).

By constrast, building the associated gravitational waveform with a
frequency-domain code is rather cumbersome.  One must build the
Fourier expansion of the waves from many coefficients, and accurately
sum them to produce the wave at any moment of time.  The benefit of
each harmonic being independent of all others is lost.  In the time
domain, building the waveform is automatic --- modulo two time
derivatives, the waveform {\it is} the observable that the code
produces.  Given a worldline, it is straightforward to build a source
for the time-domain wave equation; one then cannot help but compute
the waveform that source generates.

We are thus confident that by using both frequency and time-domain
perturbation techniques, we can get the best of both worlds ---
letting each technique's complementary strengths shine to build EMRI
waveforms that are as accurate as possible, in the context of the
adiabatic approximation.

\subsection{This paper}

Key to the success of the hybrid approach is the development of fast,
accurate codes for both frequency and time-domain approaches to black
hole perturbation theory.  First results from a frequency-domain code
which can handle generic orbits have recently been presented
{\cite{dh06}}, and the last major formal step (understanding the
adiabatic evolution of Carter's constant $Q$ due to GW emission) is
essentially in hand {\cite{dfh05, sthn05, sthgn06}}. The
frequency-domain side of this program is thus in a good state.  Our
goal now is to develop time-domain tools sufficiently robust and
generic to handle the case of interest.

The major difficulty in building a time-domain perturbation code is
the source term, representing the smaller member of the binary which
perturbs the large black hole's spacetime.  In the frequency domain,
the small body is usually approximated as having zero spatial extent,
and can be represented using delta functions (and their derivatives).
One then constructs a Green's function from solutions of the
source-free perturbation equation and integrates over the source.
Thanks to the delta nature of the source in this representation, this
integral can be done analytically.  This trick cannot be done in the
time domain --- one must choose a functional form of the source which
can be represented on a finite difference grid.  The challenge is to
pick a representation that accurately captures the very narrow spatial
extent of the source, but is sufficiently smooth that the source does
not seed excessive amounts of numerical error.  This is particularly
difficult for sources representing highly dynamic, generic Kerr black
hole orbits in which the source rapidly moves across the grid.

Much recent success in this approach has come from representing the
source as a truncated, narrow Gaussian {\cite{akp03}}.  Khanna
{\cite{k04}} and Burko and Khanna {\cite{bk07}} have so far examined
some orbit classes (equatorial orbits, both circular and eccentric)
and found that they can quickly and robustly generate waveforms from
orbits around Kerr black holes.  As a diagnostic of this technique,
they compute the flux of energy carried by this radiation and find
agreement with pre-existing frequency-domain calculations at the few
percent level.

An interesting recent development is the use of finite element
techniques to represent time-domain sources.  Such methods are tailor
made for resolving problems with multiple lengthscales, and as such
may be ideal for the EMRI problem.  Sopuerta and Laguna {\cite{sl06}}
have found that a finite element code makes it possible to represent
the source with amazing accuracy --- agreement with frequency-domain
calculations at the few hundredths of a percent level seems common.
To date, they have only examined binaries in which the larger black
hole is non-rotating, but they argue convincingly
{\cite{sopuerta_lisa}} that the difficulties required to model Kerr
perturbations should not be terribly difficult to surmount.  These
techniques are an extremely promising direction that is sure to
develop extensively in the next several years.
 
Our goal in this paper is to develop another representation of the
source term that is simpler (and concomitantly less accurate) than
finite element methods, but that is developed somewhat more
systematically than the truncated Gaussian.  The key ingredient of
this approach is an extension of a finite impulse representation of
the Dirac delta function {\cite{ett05,te04}}.  In essence, one writes
the discrete delta as a series of spikes on the finite difference
grid, with the largest spike centered at the argument of the delta,
and with the spikes rapidly falling off away from this center.  One
chooses the magnitude of the spikes such that the delta function's
integral properties are preserved, particularly the rule that
\begin{equation}
\int dx\,f(x)\delta(x - x_0) = f(x_0)\;.
\end{equation}
The discrete delta described in Ref.\ {\cite{ett05}} allows one to
make a tradeoff between localization and smoothness --- one can smear
the delta over $k$ points, choosing $k$ to be small if source
sharpness is the key property needed, or allowing $k$ to expand if too
much sharpness causes numerical problems.  This representation
introduces a kind of optimization parameter which one can engineer as
needed to find the best compromise between smoothness and
localization.

We extend the finite impulse representation of the delta described in
{\cite{ett05,te04}} in two important ways.  First, the source term of
the Teukolsky equation requires not just the delta, but also the
delta's first and second derivatives.  We therefore generalize this
procedure to develop discrete delta derivatives.  If the delta is
represented by $k$ points, then both derivatives will require $k + 2$
points.  The guiding principle of this extension is again the notion
that the integral properties of these functions must be preserved:
\begin{eqnarray}
\int dx\,f(x)\delta'(x - x_0) = -f'(x_0)\;,
\nonumber\\
\int dx\,f(x)\delta''(x - x_0) = f''(x_0)\;.
\end{eqnarray}
(Here, prime means $d/dx$.)

If the discrete delta function does not lie precisely on a grid point,
then one must use interpolation to appropriately weight impulse
functions from the neighboring grid points.  Our second extension of
Ref.\ {\cite{ett05}} is to introduce higher order interpolation
(cubic) which offers another way to trade smoothness for localization.
This is particularly valuable when (as in our application) the source
is coupled to a wave equation.

We test this representation by developing a new time-domain Teukolsky
equation solver which uses this form of the delta for its source (the
``$\delta$-code'') and comparing to a well-established code (see,
e.g., \cite{akp03,k04,bk07}) which uses a truncated Gaussian (the
``G-code'').  The G-code has been described in detail in a previous
publication {\cite{akp03}}; for the purpose of this paper, the most
salient feature of this code is how it represents the source term.
The G-code begins with the following approximation to the Dirac delta
function:
\begin{equation}
\delta[x - x(t)] \simeq \frac{1}{\sqrt{2\pi}\sigma}\exp\left(-\frac{[x -
x(t)]^2}{2\sigma^2}\right)\;.
\label{eq:gauss_delta}
\end{equation}
[Cf.\ Ref.\ {\cite{akp03}}, Eq.\ (19).] The width $\sigma$ is chosen
to be small enough that this delta only spreads across a few grid
zones.  The Teukolsky equation source is then built from this Gaussian
representation and its derivatives.

The $\delta$-code by contrast uses the representation described in
detail in the following sections of this paper --- a representation
that is discrete by design, rather than a discretization of a
continuous delta approximation.  The principle advantage of this form
seems to be that it makes it possible to rigorously enforce integral
identities involving the delta {\it plus} its derivatives.

There are a few other minor differences between these two codes, which
are artifacts of the codes' independent developments.  Chief among
these differences are the use of slightly different axial coordinates
(the G-code uses the usual Boyer-Lindquist coordinate $\phi$; the
$\delta$-code follows Ref.\ {\cite{klpa97}} and uses a coordinate
$\tilde\phi$ defined in Sec.\ {\ref{sec:h_teukeqn}}), and the use of
slightly different fundamental ``fields'' (i.e., slightly different
representations of the Weyl curvature scalar $\psi_4$ which the
Teukolsky equation governs).  There are also some differences in the
way the two codes implement boundary conditions.  We present a
detailed comparison of the results from the two codes in Sec.\
{\ref{sec:results}}.  It's worth pointing out that that we also have
taken the G-code and replaced its source term with that used by the
$\delta$-code.  This exercise confirmed all of the results we obtained
with the $\delta$-code, demonstrating that these minor differences had
no impact on our results.
 
As a proof-of-principle check of this idea's validity, we restrict our
present analysis to circular, equatorial orbits.  The results from
both codes are then compared against frequency-domain results.  Flux
of energy carried by gravitational waves is a very useful benchmark
with which to diagnose a perturbation theory code's accuracy
(especially for very simple orbits when averaging is easy both in time
and frequency domains).  In all cases, we find (after some
experimentation to optimize our discrete delta) that this new source
form is more accurate (typically by factors of $2 - 5$) and faster
(often by factors of about 10) than the truncated Gaussian.  For our
purpose, it appears that this form of source function will be very
well-suited to serve as the core of the time-domain portion of our
hybrid approach to EMRI waveforms.

Future papers in this series will then apply this technique to flesh
out the hybrid approach.  Our first followup will examine how well
this source works for highly dynamical trajectories --- generic
(inclined and eccentric) geodesic orbits, plunging orbits, and
non-geodesic trajectories (standing in for orbits that evolve due to
radiation reaction).  Early results from this analysis indicate that
the discrete delta source term handles such orbits very robustly,
validating earlier results for eccentric equatorial orbits
{\cite{bk07}}; work is in progress to extend this to generic orbits.
We will then begin developing hybrid EMRI waveforms in earnest, using
frequency-domain tools to compute the effects of radiation reaction,
building an inspiral worldline from those effects, and finally
computing the waveform with our time-domain code.

\subsection{Organization of this paper}

The remainder of this paper is organized as follows.  Section
{\ref{sec:evol}} reviews how one solves the Teukolsky equation in the
time domain, introducing the equation itself, specializing to the form
that we use for our calculations, and showing how to extract waveforms
and fluxes from its solutions.  We first review in Sec.\
{\ref{sec:h_teukeqn}} how one solves for the homogeneous (source-free)
form of the Teukolsky equation, an important first step to developing
a robust solver for the sourced case.  We follow very closely the
procedure laid out in Ref.\ {\cite{klpa97}}; this section is thus
largely a review and summary of that paper (with a few minor
corrections noted).  Section {\ref{sec:source}} then describes in
detail the form of the source term that applies when perturbations
arise from an orbiting body.

The need to model this source using a delta function motivates Sec.\
{\ref{sec:delta}}, our model for a discrete delta and its derivatives.
This section presents the key new idea of this paper.  After
describing the basic idea behind our discrete delta, we first present
in some detail (Sec.\ {\ref{sec:simple}}) an extremely simple
two-point discrete delta function.  This illustrates the concepts and
principles of this approach.  We then generalize this idea to a
multiple point delta in Sec.\ {\ref{sec:npoint}}, and then show how to
smooth things with higher order interpolation in Sec.\
{\ref{sec:cubic}}.  Some preliminary issues related to the convergence
of quantities computed using the discrete delta are introduced in
Sec.\ {\ref{sec:delta_convergence}}.

We test this delta representation in Sec.\ {\ref{sec:results}},
examining how the various methods we develop work at describing the
Teukolsky source function.  Section {\ref{sec:compare}} first compares
the different discrete delta functions with each other, demonstrating
how the different approaches change the quality and accuracy of our
results.  Based on this analysis, we choose to use the high order
(cubic) delta described in Sec.\ {\ref{sec:cubic}} in the remainder of
our work.  We then examine the convergence of our code, demonstrating
second-order convergence in Sec.\ {\ref{sec:code_convergence}}.
Finally, in Sec.\ {\ref{sec:compare_disc_gauss}} we compare the
discrete delta with the Gaussian source function, demonstrating
explicitly how this new representation improves both the code's speed
and accuracy.

Our benchmark for evaluating our results is to compare the energy flux
carried by the system's emitted gravitational waves to results
obtained using a frequency-domain code {\cite{h2000,dh06}}.  This
operation requires us to extract these waves at a particular finite
radius.  Section {\ref{sec:rad_extract}} examines the dependence of
these fluxes as a function of extraction radius, and finds that they
are very well fit by a simple power law.  Using this law, we can
easily extrapolate our results to very large radius; doing so greatly
improves agreement with frequency-domain results, typically indicating
that our errors are significantly smaller than $1\%$ for a large
fraction of parameter space.

Concluding discussion in given in Sec.\ {\ref{sec:summary}}.  Besides
summarizing the major findings of this analysis, we discuss in some
detail future projects to which we intend to apply this new
computational technology.
 
\section{Numerical implementation of the Teukolsky equation in the time domain}
\label{sec:evol}

Here we describe the evolution algorithm used in the $\delta$-Code,
built using a two step Lax-Wendroff algorithm.  Our notation and
approach closely follow that used in ~\cite{klpa97}; some of this
section therefore can be considered a summary of that paper.  All
details related to the G-Code were described in ~\cite{akp03}.

Teukolsky derived a master equation that describes perturbations due
to scalar, vector and tensor fields in the vicinity of Kerr black
holes in~\cite{ t73,TPRL}. In Boyer-Lindquist coordinates, this
equation reads
\begin{eqnarray}
\label{teuk0}
&&
-\left[\frac{(r^2 + a^2)^2 }{\Delta}-a^2\sin^2\theta\right]
         \partial_{tt}\Psi
-\frac{4 M a r}{\Delta}
         \partial_{t\phi}\Psi \nonumber \\
&&- 2s\left[r-\frac{M(r^2-a^2)}{\Delta}+ia\cos\theta\right]
         \partial_t\Psi\nonumber\\  
&&
+\,\Delta^{-s}\partial_r\left(\Delta^{s+1}\partial_r\Psi\right)
+\frac{1}{\sin\theta}\partial_\theta
\left(\sin\theta\partial_\theta\Psi\right)+\nonumber\\
&& \left[\frac{1}{\sin^2\theta}-\frac{a^2}{\Delta}\right] 
         \partial_{\phi\phi}\Psi\nonumber\\
&&+\, 2s \left[\frac{a (r-M)}{\Delta} + \frac{i \cos\theta}{\sin^2\theta}
\right] \partial_\phi\Psi\nonumber\\
&&- \left(s^2 \cot^2\theta - s \right) \Psi = -4\pi\left(r^2+a^2\cos^2\theta\right)T  ,
\end{eqnarray}
where $M$ is the mass of the black hole, $a$ its angular momentum per
unit mass, $\Delta = r^2 - 2 M r + a^2 = (r - r_+)(r-r_-), \, r_\pm
=M\pm\sqrt{M^2-a^2} \,$ and $s$ is the ``spin weight'' of the
field. The $s = \pm 2$ versions of these equations describe
perturbations to the Weyl curvature tensor, in particular the
radiative degrees of freedom $\psi_0$ and $\psi_4$.  That is, $\Psi =
\psi_0$ for $s = +2$, and $\Psi = \rho^{-4}\psi_4$ for $s=-2$, with
$\rho=-1/(r-ia\cos\theta)$.  The $T$ in the RHS of this equation
depends on the details of the perturbing source. It is here that the
Dirac delta function and its derivatives enter. A discussion of $T$ is
postponed to the latter half of this section, after we discuss the
numerical evolution of the homogeneous Teukolsky equation.

Gravitational waves, $h_+$ and $h_\times$ as well as the energy flux
$dE/dt$ \cite{fluxform1,fluxform2}, can be obtained far away from the
system by using $s=-2$ in Eq.\ (\ref{teuk0}) and then identifying
\begin{eqnarray}
\label{eq:hphc}
\psi_4 & = & \frac{1}{2}\left(\frac{\partial^2h_{+}}{\partial t^2} - i \frac{\partial^2h_{\times}}{\partial t^2} \right)\; ,
\end{eqnarray}
\begin{eqnarray}
\frac{dE}{dt} & = &\lim_{r\rightarrow\infty}\left[\frac{1}{4\pi r^6}\int_\theta\int_\phi\sin\theta\;d\theta\;d\phi\left|\int_{-\infty}^td\tilde{t}\Psi(\tilde{t},r,\theta,\phi)\right|^2\right]\nonumber\\
				& = & \lim_{r\rightarrow\infty}\left[\frac{1}{2}\int_\theta \sin\theta\;d\theta\left|\int_{-\infty}^td\tilde{t}\Phi(\tilde{t},r,\theta)\right|^2\right]\; .
\end{eqnarray}
The $\theta$ and $\phi$ directions are taken with respect to the black
hole's spin axis; the function $\Phi(t,r,\theta)$ is a reweighting of
the field $\Psi$ which we define precisely in Eq.\ (\ref{eq:psiphi})
below.

\subsection{Homogeneous Teukolsky equation}
\label{sec:h_teukeqn}

Reference~\cite{klpa97} demonstrated stable numerical evolution of
(\ref{teuk0}) for $s=-2$. The $\delta$-code has been built using the
algorithm presented in~\cite{klpa97}, after accounting for some
typographical errors, which are also discussed in~\cite{avaloslousto}.
The contents of this section are largely review of the results
presented in~\cite{klpa97}; as such, our discussion is particularly
brief here.

Our code uses the tortoise coordinate $r^*$ in the radial direction,
and azimuthal coordinate $\tilde{\phi}$; these coordinates are related
to the usual Boyer Lindquist quantities by
\begin{eqnarray}
dr^* &=& \frac{r^2+a^2}{\Delta}dr \\
\Rightarrow r^* &=& r + \frac{2Mr_+}{r_+-r_-}\ln\frac{r-r_+}{2M} \nonumber\\
				&  &- \frac{2Mr_-}{r_+-r_-}\ln\frac{r-r_-}{2M} \; ,
\end{eqnarray}
and
\begin{eqnarray}
d\tilde{\phi} &=& d\phi + \frac{a}{\Delta}dr\\
\Rightarrow \tilde{\phi} &=& \phi + \frac{a}{r_+-r_-}\ln\frac{r-r_+}{r-r_-}\; .
\end{eqnarray}  
Following \cite{klpa97}, we factor out the azimuthal dependence and use the ansatz,
\begin{eqnarray}
\label{eq:psiphi}
\Psi(t,r^*,\theta,\tilde{\phi}) &=& e^{im\tilde{\phi}} r^3 \Phi(t,r^*,\theta) .
\end{eqnarray}
Defining
\begin{eqnarray}
\Pi &\equiv& \partial_t{\Phi} + b \, \partial_{r^*}\Phi \; , \\
b & \equiv &
\frac { {r}^{2}+{a}^{2}}
      { \Sigma} \; , 
\end{eqnarray}
and
\begin{eqnarray}
\Sigma^2 &\equiv &  (r^2+a^2)^2-a^2\,\Delta\,\sin^2\theta
\; 
\label{pi_eq}
\end{eqnarray} 
allows the Teukolsky equation to be rewritten as
\begin{eqnarray}
\label{eq:evln}
\partial_t \mbox{\boldmath{$u$}} + \mbox{\boldmath{$M$}} \partial_{r*}\mbox{\boldmath{$u$}} + \mbox{\boldmath{$Lu$}} + \mbox{\boldmath{$Au$}} =  \mbox{\boldmath{$T$}} ,
\end{eqnarray}
where 
\begin{equation}
\mbox{\boldmath{$u$}}\equiv\{\Phi_R,\Phi_I,\Pi_R,\Pi_I\}
\end{equation}
is the solution vector. The subscripts $R$ and $I$ refer to the real
and imaginary parts respectively. (Note that the Teukolsky function
$\Psi$ is a complex quantity.) The matrices {\boldmath{$M$}},
{\boldmath{$A$}} and {\boldmath{$L$}} are
\begin{equation}
\mbox{\boldmath{$M$}} \equiv \left(\begin{matrix}
                    b  &   0   &  0     &  0 \cr
                    0  &   b   &  0     &  0 \cr
                    m_{31}  &   m_{32}  & -b  &  0 \cr
                    -m_{32}  &   m_{31} &  0  & -b \cr
                \end{matrix}\right) \; ,
\label{m_matrix}
\end{equation}
\begin{equation}
\mbox{\boldmath{$A$}} \equiv \left(\begin{matrix}
                    0  &   0   &  -1  &  0 \cr
                    0  &   0   &  0  &  -1 \cr
                    a_{31} & a_{32} & a_{33} & a_{34} \cr
                    -a_{32} & a_{31} & -a_{34} & a_{33} \cr
                \end{matrix}\right) \; ,
\label{a_matrix}
\end{equation}
and
\begin{equation}
 \mbox{\boldmath{$L$}} \equiv \left(\begin{matrix}
                    0  &   0   &  0  &  0 \cr
                    0  &   0   &  0  &  0 \cr
                    l_{31}  &   0   &  0  &  0 \cr
                    0  &   l_{31}   &  0  &  0 \cr
                \end{matrix}\right)\;,
\label{l_matrix}
\end{equation}
where 
\begin{eqnarray}
\label{eq:coeffs}
m_{31} & = & -b c_1  + b\partial_{r^*}b + c_2  \; ,\\
m_{32} & = & b c_3 + 2a m (r^2 + a^2)/\Sigma^2  \; ,\\
a_{31} & = & \Delta\frac{m^2 + 2\cos\theta s m + \cos^2\theta s^2 - \sin^2\theta s}{\Sigma^2 \sin^2\theta}
\nonumber\\
		&  &- 6\Delta \frac{a^2 + r\left(r\left(s+2\right) - M\left(s+3\right)\right)}{r^2 \Sigma^2}\; ,\\
a_{32} & = & \frac{4M(r-1)s m a M + (6 a m \Delta)/r}{\Sigma^2}\; ,\\
a_{33} & = & c_1\; ,\\
a_{34} & = & -c_3 \; ,\\
l_{31} & = & -\frac{\Delta}{\Sigma^2}\frac{\partial^2}{\partial\theta^2} - \cot\theta\frac{\Delta}{\Sigma^2}\frac{\partial}{\partial\theta} \; ,\\
c_1 & = & 2 s  (-3 Mr^2 + Ma^2 + r^3 + r a^2)/\Sigma^2 \; ,\\
c_2 & = & -2\frac{r\Delta\left(1+s\right) - (a^2 - r^2) M s}{\Sigma^2} - \frac{6\Delta b}{r\Sigma}\; ,\\
c_3 & = & 2 a (2 r M m + \Delta s\cos\theta)/\Sigma^2\; .
\end{eqnarray}
The equations above have been written such that the typographical
errors in \cite{klpa97}'s $a_{31},a_{32},a_{34}$ and $c_2$ are
obvious. It turns out that the coefficients listed in \cite{klpa97}
are correct when the ansatz $\Psi(t,r^*,\theta,\tilde{\phi}) =
e^{im\tilde{\phi}}\Phi(t,r^*,\theta)$ is used. {\boldmath{$T$}} is a
quantity contructed from the source term and is discussed in the
latter half of this section.

Rewriting Eq.\ (\ref{eq:evln}) as 
\begin{equation}
\partial_t \mbox{\boldmath{$u$}} + \mbox{\boldmath{$D$}}
\partial_{r^*} \mbox{\boldmath{$u$}}
=  \mbox{\boldmath{$S$}}\; , 
\label{new_teu2}
\end{equation}
where
 
\begin{equation}
 \mbox{\boldmath{$D$}} \equiv \left(\begin{matrix}
                    b &   0   &  0  &  0 \cr
                    0  &   b   &  0  &  0 \cr
                    0  &   0   &  -b  &  0 \cr
                    0  &   0   &  0  &  -b \cr
                \end{matrix}\right),
\label{d_matrix}
\end{equation}

\begin{equation}
\mbox{\boldmath{$S$}} =\mbox{\boldmath{$T$}} -(\mbox{\boldmath{$M$}} - \mbox{\boldmath{$D$}})
\partial_{r^*}\mbox{\boldmath{$u$}}
- \mbox{\boldmath{$L$}}\mbox{\boldmath{$u$}} 
- \mbox{\boldmath{$A$}}\mbox{\boldmath{$u$}},
\end{equation}
and subjecting it to Lax-Wendroff iterations produces stable
time-evolutions. Each Lax-Wendroff iteration consists of two steps. In
the first step, the solution vector between grid points is obtained
from
\begin{eqnarray}
\mbox{\boldmath{$u$}}^{n+1/2}_{i+1/2} &=& 
\frac{1}{2} \left( \mbox{\boldmath{$u$}}^{n}_{i+1}
                  +\mbox{\boldmath{$u$}}^{n}_{i}\right)
- \\
&  &\frac{\delta t}{2}\,\left[\frac{1}{\delta r^*} \mbox{\boldmath{$D$}}^{n}_{i+1/2}
  \left(\mbox{\boldmath{$u$}}^{n}_{i+1}
                  -\mbox{\boldmath{$u$}}^{n}_{i}\right)
- \mbox{\boldmath{$S$}}^{n}_{i+1/2} \right] \; .\nonumber
\end{eqnarray}
This is used to compute the solution vector at the next time step,
\begin{equation}
\mbox{\boldmath{$u$}}^{n+1}_{i} = 
\mbox{\boldmath{$u$}}^{n}_{i}
- \delta t\, \left[\frac{1}{\delta r^*} \mbox{\boldmath{$D$}}^{n+1/2}_{i}
  \left(\mbox{\boldmath{$u$}}^{n+1/2}_{i+1/2}
                  -\mbox{\boldmath{$u$}}^{n+1/2}_{i-1/2}\right)
- \mbox{\boldmath{$S$}}^{n+1/2}_{i} \right] \, .
\end{equation}
The angular subscripts are dropped here for clarity. All angular
derivatives were computed using second order centered finite
difference expressions. Notice that the matrices {\boldmath{$D$}},
{\boldmath{$A$}} and {\boldmath{$M$}} are time independent. In
addition, the time stepping must satisfy the Courant-Friedrichs-Lewy
condition~\cite{klpa97}, $ \delta t \le \mbox{max}\{\,\delta r^*, \,
5M\, \delta\theta\, \}$, where $\delta t$ is the time
step\footnote{Conducting a von Neumann local stability analysis on all
the points of our numerical grid yields that this condition is
sufficient for stable evolutions. See reference~\cite{klpa97} for more
details. }.

Following \cite{klpa97}, we set $\Phi$ and $\Pi$ to zero on the inner
and outer radial boundaries. While the aymptotic behavior
\begin{equation}
\lim_{r^*\rightarrow-\infty} |\Psi| \propto \Delta^{-s} 
\end{equation}
makes this condition reasonably accurate at the inner boundary, it is
clearly unphysical at the outer boundary.  By placing our outer
boundary sufficiently far, error due to our outer boundary condition
can be made unimportant; reflections from the outer boundary have no
important impact on our results. Symmetry of the spheroidal harmonics
is used to determine the angular boundary conditions. For even $|m|$
modes, we have $\partial_\theta\Phi =0$ at $\theta = 0,\pi$. On the
other hand, $\Phi =0$ at $\theta = 0,\pi$ for modes of odd $|m|$.

\begin{figure*}[htb]
\begin{center}
\includegraphics[height = 100mm]{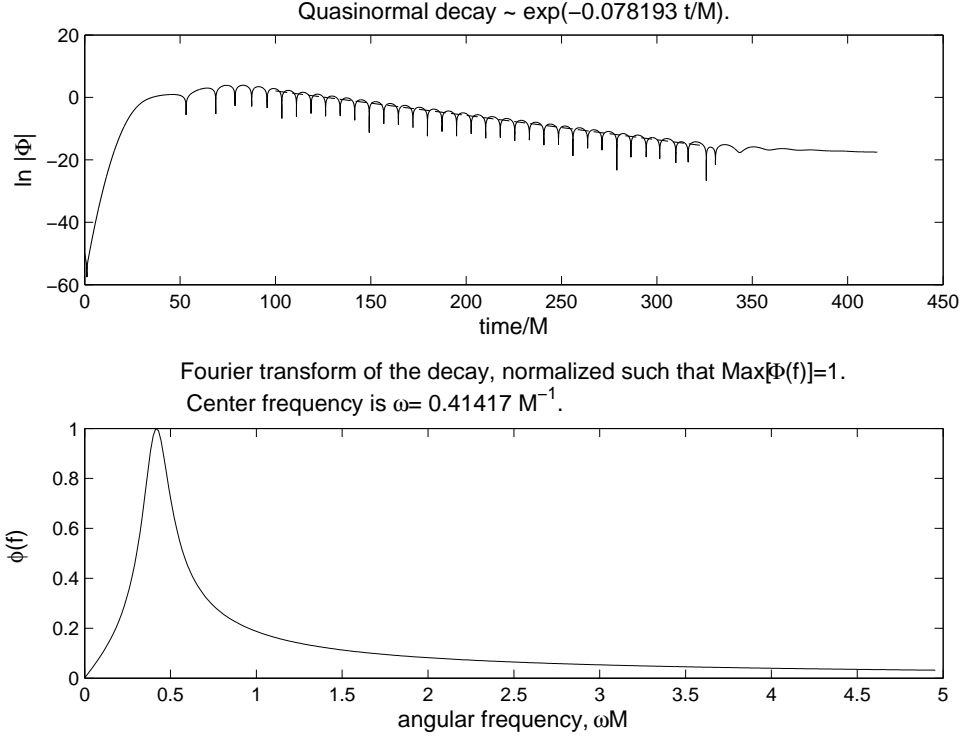}
\caption{Illustrations of quasi-normal ringing for a black hole with
$a/M = 0.9$; the $l = 2$, $m = 0$ mode is shown here.  Top panel:
Evolution of the magnitude of the Teukolsky function, extracted at $r
= 20 M$, $\theta = \pi/2$. We plot the time evolution of $\ln |\Phi|$
at this position.  Overplotted on this curve (dashed line) is a
function $\propto \exp(-0.078193 t/M)$, demonstrating that we recover
the expected decay law with a damping time $\tau = 12.789M$.  Bottom
panel: Magnitude of the Fourier transform of $\Phi(t)$.  Notice that
it peaks at $\omega = 0.41417/M$.  These results for $\omega$ and
$\tau$ are in excellent agreement with the expected values of $(\omega
M,M/\tau) = (0.41,-0.078)$ from Ref.~\cite{k91} for quasi-normal
ringing of the $l = 2$, $m = 0$ mode for $a = 0.9$.\label{qnm}}
\end{center}
\end{figure*}

As a test of our evolution equation, we have examined source-free
field evolution (setting {\boldmath{$T$}} $= 0$) for a variety of
initial data, in particular comparing extensively with the results of
\cite{klpa97}.  As an example of one of our tests, we take initial
data corresponding to an ingoing, narrow Gaussian pulse.  This data
perturbs the black hole, causing it to ring down according to its
characteristic quasi-normal frequencies.  We find extremely good
agreement in mode amplitude and evolution (typically $\sim 1\%$ error)
with results from \cite{klpa97}.  Figure \ref{qnm} shows the result of
such a test, illustrating the quasi-normal ringing and power law tail
for the $l = 2$, $m = 0$ mode of a black hole with spin parameter $a =
0.9$.

\subsection{The source term}
\label{sec:source}

We now consider the source term, $T$, of Eq. (2.1).  It is given by
\begin{eqnarray}
\label{eq:T4}
T & = & 2\rho^{-4}T_{4}\;,\\
\label{eq:sourcedef}
T_4 & = & (\tilde\Delta+3\gamma-\bar\gamma+4\mu+\bar\mu)(\tilde\Delta+2\gamma-2\bar\gamma+\bar\mu)T_{\bar m\bar m}\nonumber\\
	& & - ( \tilde\Delta+3\gamma-\bar\gamma+4\mu+\bar\mu)(\bar\delta-2\bar\tau+2\alpha)T_{n\bar m}\nonumber\\
	& & + (\bar\delta-\bar\tau+\bar\beta+3\alpha+4\pi)(\bar\delta-\bar\tau+2\bar\beta+2\alpha)T_{nn}\nonumber\\
	& & -(\bar\delta-\bar\tau+\bar\beta+4\pi)( \tilde\Delta+2\gamma+2\bar\mu)T_{n\bar m} \;.
\label{eq:T_4def}
\end{eqnarray}
 
Reference~\cite{ t73} provides definitions for the various quantities
which appear in Eqs.\ (\ref{eq:sourcedef}) and (\ref{eq:T_4def}).  Of
particular importance are the quantities $T_{nn}$, $T_{n\bar m}$, and
$T_{\bar m\bar m}$, given by contracting the stress energy tensor for
the orbiting body with the Newman-Penrose null-tetrad legs $n^\mu$ and
$\bar m^\mu$:
\begin{eqnarray}
n^\mu &\doteq& \left(\frac{\rho^{2}\left(r^2 + a^2\right)}{2},
-\frac{\Delta \rho^{2}}{2}, 0, \frac{a\rho^{2}}{2}\right)\;,
\label{eq:NP_n}
\\
\bar m^\mu &\doteq& \frac{1}{\sqrt{2}(r - i a\cos\theta)} \nonumber\\
	& & \left(-ia\sin\theta, 0, 1, -i\csc\theta\right)\;,
\label{eq:NP_mbar}
\end{eqnarray}
where $\Delta = r^2 - 2Mr + a^2$ and $
\rho^{-2} = r^2 + a^2\cos^2\theta$.
 
Also of great importance here are the Newman-Penrose operators
$\tilde\Delta$ and $\bar\delta$:
\begin{eqnarray}
\tilde\Delta &=& n^\mu \frac{d}{dx^\mu}
\nonumber\\
&=& \frac{\rho^{2}\left(r^2 + a^2\right)}{2}\frac{d}{dt} - \frac{\rho^{2}\Delta}{2}
\frac{d}{dr} + \frac{aim\rho^{2}}{2}
\nonumber\\
&\equiv& \Delta_t + \Delta_r + \Delta_\phi\;;
\label{eq:Delta_operator}
\\
\bar\delta &=& {\bar m}^\mu \frac{d}{dx^\mu}
\nonumber\\
&=& -\frac{ia\sin\theta\rho^{2}(r + ia\cos\theta)}{\sqrt{2}}\frac{d}{dt} \nonumber\\
& & + \frac{(r + ia\cos\theta)\rho^{2}}{\sqrt{2}} \frac{d}{d\theta} + \frac{m\rho^{2}(r + ia\cos\theta)}{\sqrt{2}\sin\theta}
\nonumber\\
&\equiv& \bar\delta_t + \bar\delta_\theta + \bar\delta_\phi\;.
\label{eq:delta_star_operator}
\end{eqnarray}
(The operator $\tilde{\Delta}$ is normally written without the tilde;
we have added it here to avoid confusion with $\Delta = r^2 - 2Mr +
a^2$.)

To proceed, we next must analyze the stress energy tensor describing
the small body.  A point body of mass $\mu$ disturbing the Kerr
spacetime is given by
\begin{equation}
T_{\mu\nu} = \frac{\mu u_\mu u_\nu\rho^{2}}{u^t\sin\theta}
\delta[r - R(t)]\delta[\theta - \Theta(t)]\delta[\phi - \Phi(t)]\;,
\label{eq:Tmunu}
\end{equation}
where $u^\mu = dx^\mu/d\tau$, and where $[R(t),\Theta(t),\Phi(t)]$
describe the Boyer-Lindquist coordinate worldline of the small body.
Due to axial symmetry of the Kerr spacetime, the only $\phi$
dependence in the stress-energy tensor comes from $\delta[\phi -
\Phi(t)]$.  Writing
\begin{equation}
T_{\mu\nu} = \sum_m T^m_{\mu\nu} e^{im\phi}
\label{eq:Tmunu_expand}
\end{equation}
and using the fact that
\begin{equation}
\delta[\phi - \Phi(t)] = \frac{1}{2\pi} \sum_m e^{im[\phi - \Phi(t)]}\;,
\end{equation}
we find
\begin{equation}
T^m_{\mu\nu} = \frac{\mu u_\mu u_\nu\rho^{2}}{u^t\sin\theta}
\delta[r - R(t)]\delta[\theta - \Theta(t)]e^{-im\Phi(t)}\;.
\label{eq:T_m_munu.}
\end{equation}

Using this expansion for $T_{\mu\nu}$, it is a simple matter to
construct $T_{nn} = n^\mu n^\nu T_{\mu\nu}$, $T_{n\bar m} = n^\mu \bar
m^\nu T_{\mu\nu}$, and $T_{\bar m\bar m} = \bar m^\mu \bar m^\nu
T_{\mu\nu}$.  We then insert these terms into Eqs.\
(\ref{eq:sourcedef}).  Using the chain rule repeatedly leaves us with
a (rather complicated) expression involving radial and $\theta$
derivatives of Dirac delta functions.  We thus face the task of
representing the delta function and its derivatives accurately on a
numerical grid.  This is the major innovation of this paper, and is
discussed in detail in the following section.

It should be noted at this point that, since the Teukolsky equation is
most naturally written in terms of the tortoise coordinate $r^*$, we
must describe the radial behavior of the source term in $r^*$ as well.
To this end, we replace the radial delta function and all radial
derivatives as follows:
\begin{eqnarray}
\delta[r - R(t)] &=& \frac{\delta[r^* - R^*(t)]}{|dr/dr^*|}\;,
\label{eq:delta_r*}
\\
\frac{d}{dr} &=& \frac{r^2 + a^2}{\Delta}\frac{d}{dr^*}\;.
\label{eq:r*_deriv}
\end{eqnarray}

Finally, in our numerical implementation, we define the vector
{\boldmath{$T$}} appearing in Eq. (2.7) as
\begin{equation}
\mbox{\boldmath{$T$}} \doteq \left[0,0,{\rm Re}(\hat T),{\rm Im}(\hat T)\right]\;,
\label{eq:Tvec_def}
\end{equation}
where
\begin{eqnarray}
\hat{T} & = & \frac{4 \pi \Delta \left(r^2+a^2\mbox{cos}^2{\theta}\right)}{r^3\left[\left(r^2 + a^2\right)^2 - a^2\Delta\mbox{sin}^2\theta\right]}\times \\
		&  & \exp\left(-i m  \frac{a}{r_+ - r_-}\ln\left[\frac{r-r_+}{r-r_-}\right]\right)\left[2\rho^{-4}T_4\right]\nonumber\;.
\end{eqnarray}
The exponential factor in this expression corrects for the fact that
the evolution code uses the azimuthal variable $\tilde\phi$, but the
source term is expanded in $\phi$.

\section{The discrete delta function and its derivitatives}
\label{sec:delta}

As pointed out in the previous section, the Dirac delta function
enters the Teukolsky equation because we approximate the perturbing
mass by a point particle.  By its definition as an integrable
singularity, the delta function is very difficult to represent on a
finite difference grid.  The best we can hope to do is to develop a
model function that captures its most important features, particularly
localization to a very small spatial region, as well as integrability
and derivative properties. The following three subsections describe
the model for the delta function we have developed.  We first describe
a very basic model that demonstrates how to satisfy our criteria in
Sec.\ {\ref{sec:simple}}.  In Sec.\ {\ref{sec:npoint}} and
{\ref{sec:cubic}}, we then refine this basic model.  These refinements
have been found to improve the overall accuracy of the code.

The discrete delta function approach we use is inspired by the work
presented in Ref.\ \cite{ett05}.  Our technique can be considered an
extension of that used in \cite{ett05}; in particular, they do not
develop delta function derivatives, nor do they implement all the
refinements discussed in {\ref{sec:npoint}} and {\ref{sec:cubic}}.  
Nonetheless, Ref.\ \cite{ett05} played an extremely important role in
developing the foundations of our work.

Before turning to a detailed discussion of our techniques for modeling
the delta function on a numerical grid, we first mention some general
considerations pertaining to delta functions on a finite difference
grid.   For concreteness, consider a function and delta combination,
$f(x)\delta(x - \alpha)$.   The function $f(x)$ is taken to be known,
and can be calculated for any $x$.   For the sake of argument, let the
delta be modelled by two coefficients $\delta_k$ and $\delta_{k+1}$ on
grid points $x_k$ and $x_{k+1}$ respectively; the delta is taken to be
zero everywhere else.  (This in fact pertains to the form of the delta
discussed in Sec.\ {\ref{sec:simple}}.)

Now imagine integrating $f(x)\delta(x - \alpha)$ over all $x$.
Analytically, we know that this should give us $f(\alpha)$.   Numerically
integrating this on our grid gives us

\begin{eqnarray}
h\sum_{i}f(x_i)\delta(x_i-\alpha)&=&h[f(x_k)\delta_k \nonumber\\
				 &  & + f(x_{k+1})\delta_{k+1}].
\end{eqnarray}

This equation suggests that the numerical integral approximates
$f(\alpha)$ by interpolating between grid points $x_k$ and $x_{k+1}$.
If $f(x)$ is rapidly varying, this interpolation may not be accurate
enough; this is sure to be a source of error as we integrate our PDE
forward in time.

Great improvement can be achieved by enforcing the well-known identity

\begin{eqnarray}
f(x)\delta(x - \alpha) = f(\alpha)\delta(x - \alpha).
\end{eqnarray}
The numerical integral now becomes
\begin{eqnarray}
h\sum_{i}f (\alpha )\;\delta (x_i-\alpha )
&=& h\left[f (\alpha )\delta_k + f(\alpha )\delta_{k+1}\right]
\nonumber\\
&=& f (\alpha )h\left(\delta_k + \delta_{k+1}\right)
\nonumber\\
&=& f (\alpha ) \;,
\end{eqnarray}
and the identity is preserved exactly. In the last step, we use the
discrete analog of the property
\begin{equation}
\int dx\;\delta(x-\alpha) = 1\;.
\end{equation}

Similar identities can be used on the delta function derivatives:
\begin{eqnarray}
f (x ) \delta^\prime (x-\alpha ) & = & f (\alpha ) \delta^\prime (x-\alpha ) \nonumber\\
								& & - f^\prime (\alpha ) \delta (x-\alpha ) \;,\\
f(x)\delta^{\prime\prime} (x-\alpha ) & = & f (\alpha ) \delta^{\prime\prime} (x-\alpha ) -2f^	\prime (\alpha ) \delta^\prime (x-\alpha )\nonumber\\ 
				& &+f^{\prime\prime} (\alpha ) \delta (x-\alpha )\;.
\end{eqnarray} 

We recommend using these identities as much as possible when
numerically implementing the algorithms sketched in the following
three subsections.

\subsection{A simple numerical delta function} 
\label{sec:simple}

Consider the function $\delta(x-\alpha)$, where $x_k \leq \alpha \leq
x_{k+1}$; i.e, $\alpha$ lies between two discrete grid points. Let
$h=x_{k+1}-x_k=x_k-x_{k-1}$ be the grid resolution. We use the
following integral to define the delta function:

\begin{eqnarray}
\int_{\alpha-\epsilon}^{\alpha+\epsilon} dx\;f(x)\;\delta(x-\alpha) & = & f(\alpha)\;,
\end{eqnarray}
where $\epsilon>0$ and $f\left(x\right)$ is any well behaved
function. This means that $\delta(x-\alpha)$ is zero everywhere,
except at $x=\alpha$, where it is singular. Translating this integral
to a summation, we have:

\begin{eqnarray}
\int_{\alpha-\epsilon}^{\alpha+\epsilon} dx\;f(x)\;\delta(x-\alpha) & \simeq & h\sum_{i}f\left(x_i\right)\delta_i\\
\Rightarrow f\left(\alpha\right)		& \simeq &   h\sum_{i}f\left(x_i\right)\delta_i\;,
\end{eqnarray}
where $\delta_i$ is the discrete delta defined on the grid. Since
$\alpha$ does not necessarily lie on a gridpoint, we can linearly
interpolate to find:

\begin{eqnarray}
f(\alpha) &=& \frac{f(x_{k+1}) - f(x_k)}{h}(\alpha - x_k) + f(x_k)
\nonumber\\
& & + O(h^2)\;.
\end{eqnarray}
Substituting this back into our earlier expression and comparing
coefficients, we have
\begin{eqnarray}
\delta_i & = & \frac{\alpha - x_k}{h^2} \mbox{ for } i=k+1 \nonumber\\
				 & = & \frac{x_{k+1} - \alpha}{h^2} \mbox{ for } i=k \nonumber\\
				 & = & 0 \mbox{ everywhere else } \;.
\end{eqnarray}
Notice that if $\alpha = x_k$, then $\delta_i = 1/h$ for $i = k$, but
is zero everywhere else; a similar result holds if $\alpha = x_{k+1}$.
This reproduces our intuitive notion that the delta function is zero
everywhere except at a single point, and that it integrates to unity.
We take the viewpoint that the integrability of the delta function is
its key defining property, using this rule to derive the results
presented below.   This is the approach that was used in
Ref.~\cite{ett05}.

Another approach to defining a numerical delta function, suggested
in~\cite{k04}, is to first define a step function on the grid, and
then use finite differencing to obtain the delta and its
derivatives.   The approach described above matches this proposal when
$\alpha$ lies exactly on a grid point.

We can proceed in a similar fashion to find formulae for the
derivatives. Let us define
\begin{equation}
\gamma = \frac{x_{k+1}-\alpha}{h} = 1 - \frac{\alpha-x_{k}}{h}\;.
\end{equation}
Again, we start from the defining integrals, 
\begin{eqnarray}
\int dx\;f(x)\;\delta^\prime(x-\alpha) & =& - f^\prime(\alpha) \;,\\
\int dx\;f(x)\;\delta^{\prime\prime}(x-\alpha) & = & f^{\prime\prime}(\alpha)\;.
\end{eqnarray}   
Note that a prime denotes $d/dx$. Our goal is to derive a form which
enforces these integrals in summation form:
\begin{eqnarray}
\label{delta_identity1}
h \sum_{i}\;f(x_i)\;\delta^\prime_i & \simeq & -f^\prime(\alpha) \nonumber\\
	& = & -h \sum_i f'(x_i)\delta_i \nonumber \\
	& = & -\gamma f'(x_k) \nonumber \\
	&   & - (1 - \gamma)f'(x_{k+1}) + O(h^2); 
\end{eqnarray}
\begin{eqnarray}
\label{delta_identity2}
h \sum_{i}\;f(x_i)\;\delta^{\prime\prime}_i & \simeq & f^{\prime\prime}(\alpha)\nonumber\\
	& = & h\sum_{i}\;f^{\prime\prime}(x_i)\delta_i \nonumber \\
	& = & \gamma f^{\prime\prime}(x_{k}) + \nonumber\\ 
	&  	& (1-\gamma)f^{\prime\prime}(x_{k+1})+ O\left(h^2\right).
\end{eqnarray} 
The derivatives of $f(x_k)$ are given by the finite difference formulae,
\begin{eqnarray}
\label{fd_deriv1}
f^\prime\left(x_{k}\right) & = & \frac{f\left(x_{k+1}\right)-f\left(x_{k-1}\right)}{2h} + O(h^2)\;,\\
\label{fd_deriv2}
f^{\prime\prime}\left(x_{k}\right) & = & \frac{f\left(x_{k+1}\right) - 2 f\left(x_{k}\right) + f\left(x_{k-1}\right)}{h^2} \nonumber\\
&	& + O(h^2).
\end{eqnarray}
Substitution of these approximations in (\ref{delta_identity1}) and
(\ref{delta_identity2}) and a comparison of coefficients yields for
the derivative:
\begin{eqnarray} 
\delta^\prime_i & = & \frac{\gamma}{2h^2} \mbox{ for } i = k - 1 \nonumber\;, \\
               	& = & \frac{1 - \gamma}{2h^2} \mbox{ for } i = k \nonumber \;,\\
               	& = & -\frac{\gamma}{2h^2} \mbox{ for } i = k + 1 \nonumber \;,\\
               	& = & \frac{\gamma - 1}{2h^2} \mbox{ for } i = k + 2 \nonumber \;,\\
     	          &  = & 0 \mbox{ everywhere else }\;.                             
\end{eqnarray}
For the second derivative:
\begin{eqnarray} 
\delta^{\prime\prime}_i & = & \frac{\gamma}{h^3} \mbox{ for } i = k - 1 \nonumber \;,\\
                 		& = & \frac{1 - 3\gamma}{h^3} \mbox{ for } i = k \nonumber \;,\\
                 		& = & \frac{3\gamma - 2}{h^3} \mbox{ for } i = k + 1 \nonumber\;, \\
                 		& = & \frac{1 - \gamma}{h^3} \mbox{ for } i = k + 2 \nonumber \;,\\
                 		& = & 0 \mbox{ everywhere else }\;.
\end{eqnarray} 
Notice that we need four points to represent the derivatives of the
delta function in this scheme.

\subsection{A multiple point delta function}
\label{sec:npoint}

The procedure described in Sec.\ {\ref{sec:simple}} can be extended to
represent the delta function over a larger number of points.   On the
one hand, this spreads out the delta, moving us away from our ideal of
a function that is non-zero in as small a region as possible; on the
other hand, it allows us to represent it more smoothly on our grid.
The number of points $(2n+2)$ that we use can thus be considered an
optimization parameter, allowing us to trade localization for
smoothness.   As we shall see, there is typically a value of $n$ that
represents a very good compromise.

We start off with the `linear hat' delta function defined in
\cite{ett05} and \cite{te04}
 
\begin{eqnarray}
\delta_i = \left\{ 
 \begin{array}{l l}
  \gamma_i/h & \quad \mbox{for} \left|x_i-\alpha\right| \leq \epsilon = nh \\
  0 & \quad \mbox{otherwise}\\ 
  \end{array} \;,\right.
\end{eqnarray}
where
\begin{equation}
\gamma_i = \frac{1}{n}\left(1 - \frac{|x_i - \alpha|}{nh}\right) \;,
\end{equation}
and $n$ is an integer.   Note that when $n = 1$, $\gamma_i$ reduces to
the $\gamma$ that was defined in Sec.\ {\ref{sec:simple}}.   Note also
that $\gamma_i$ is non-zero only for $i \in [k,...,k+2n-1]$(so that
there are a total of $2n$ points), and that $\alpha$ lies between the
grid points $x_{k + n - 1}$ and $x_{k + n}$.  In this labeling scheme,
$x_k$ is the smallest gridpoint where $\delta_i$ is nonzero.

Substituting this form of the delta function into our defining
integral relation,

\begin{equation}
h\sum_i f(x_i) \delta_i \simeq f(\alpha)\;,
\end{equation}
we find
\begin{equation}
\sum_i f(x_i) \gamma_i \simeq f(\alpha)\;,
\end{equation}
The quantity $\gamma_i$ is thus a weighting factor whose weight
depends on the distance of $x_i$ from $\alpha$.   Setting $f(x) = 1$,
we find the property
\begin{equation}
\sum_i \gamma_i = 1\;.
\end{equation}

Now consider the derivative of the delta function.   Our goal is again
to enforce the rule

\begin{equation}
h \sum \delta'_i f(x_i) \simeq -f'(\alpha) \simeq -h\sum\delta_i f'(x_i)\;.
\end{equation}
Inserting the finite difference formulae for the derivatives of $f$,
Eqs.   (\ref{fd_deriv1}) and (\ref{fd_deriv2}), into this relation, we
find
\begin{eqnarray}
f^\prime(\alpha) & \simeq & -\sum_{i=k}^{k+2n-1}\gamma_i \left[\frac{f\left(x_{i+1}\right) - f\left(x_{i-1}\right)}{2h}\right]\nonumber\\
	& = & -\frac{1}{2h}\left[\left(\sum_{i=k+1}^{k+2n-2}\gamma_{i-1}f(x_i)\right)\right] \nonumber\\
	&  & + \frac{1}{2h}\left[ \gamma_{k+2n-2} f(x_{k+2n-1})\right] \nonumber\\
	&  & + \frac{1}{2h}\left[ \gamma_{k+2n-1} f(x_{k+2n})\right] \nonumber\\
	& 	&  - \frac{1}{2h} \left[\left(\sum_{i=k+1}^{k+2n-2}\gamma_{i+1}f(x_i)\right)\right]\nonumber\\
	& 	&  +\frac{1}{2h}\left[-\gamma_k f(x_{k-1}) - \gamma_{k+1} f(x_{k})\right]\;. 
\end{eqnarray}
Comparing coefficients, we read off
\begin{eqnarray}
\delta^\prime_{k-1} & = & \frac{\gamma_k}{2h^2} \;,\\
\delta^\prime_k     & = & \frac{\gamma_{k+1}}{2h^2} \;,\\
\delta^\prime_{k+j} & = & -\frac{\gamma_{k+j-1}-\gamma_{k+j+1}}{2h^2} \nonumber\\
		    & 	&	\mbox{ for } j\in \left[1,2n-2\right] \;,\\
\delta^\prime_{k+2n-1} & = & -\frac{\gamma_{k+2n-2}}{2h^2} \;, \\ 
\delta^\prime_{k+2n} & = &  -\frac{\gamma_{k+2n-1}}{2h^2} \;.
\end{eqnarray} 
 
The formulas for the delta derivative coefficients can be understood
intuitively.   The $n$-point generalization approximates the delta
function as an isoceles triangle centered at $\alpha$ and sampled at
$2n$ points.   The derivative is simply the slope of this isoceles
triangle at all points, except at the center and the edges, where the
derivative is discontinuous.   The discontinuity is replaced by
coefficients that ensure the integral properties of the
derivative. The delta derivative takes a particularly simple formula
in the ``bulk'':
\begin{eqnarray}
\delta^\prime_{k+j} & =&  -\frac{\gamma_{k+j-1}-\gamma_{k+j+1}}{2h^2} \mbox{ for } j\in \left[1,2n-2\right]\nonumber\\
					& = & \frac{1}{2nh^2}\left[\frac{\left|x_{k+j-1}-\alpha\right|}{nh}-\frac{\left|x_{k+j+1}-\alpha\right|}{nh}\right]\nonumber\\
					& =  & \left\{{ 
 							\begin{array}{l l}
  							\frac{1}{2n^2h^2} & \quad \mbox{ for }  x_{k+j+1} < \alpha - h\\
 						   - \frac{1}{2n^2h^2} & \quad \mbox{ for }  x_{k+j+1} > \alpha + h\\
  							\end{array}\;.} \right.
\end{eqnarray}
Notice that the delta derivative coefficients are non-zero for $i \in
[k - 1, k +2n]$ --- one point wider in each direction than the span of
the delta on the grid.

A similar analysis can be done for the second derivatives. We start off with
\begin{eqnarray}
f^{\prime\prime}(\alpha) & \simeq & \sum_{i=k}^{k+2n-1}\gamma_i f^{\prime\prime}(x)\;,\\
h \sum_{i}\delta^{\prime\prime}f^(x_i) & = &	\sum_{i=k}^{k+2n-1}	\gamma_i \left[\frac{f\left(x_{i+1}\right) - 2f\left(x_{i}\right) + f\left(x_{i-1}\right)}{h^2}\right]\;.\nonumber\\
\end{eqnarray}
Reading off the coefficients leaves us with
\begin{eqnarray}
\delta^{\prime\prime}_{k-1} & = & \frac{\gamma_k}{h^3} \;,\\
\delta^{\prime\prime}_k & = & \frac{\gamma_{k+1} - 2 \gamma_k}{h^3}\;, \\
\delta^{\prime\prime}_{k+j} & = & \frac{\gamma_{k+j+1} - 2 \gamma_{k+j} + \gamma_{k+j-1}}{h^3} = 0 \nonumber\\
							&  & \mbox{ for } j\in \left[1,2n-2\right] \;,\\
\delta^{\prime\prime}_{k+2n-1} & = & \frac{\gamma_{k+2n-2} - 2 \gamma_{k+2n-1}}{h^3} \;, \\
\delta^{\prime\prime}_{k+2n}	& = & \frac{\gamma_{k+2n-1}}{h^3} \;.
\end{eqnarray} 

Notice that the second derivative is zero in the ``bulk'' --- it
corresponds to the second derivative of a line, with constant slope.
Like the first derivative, these coefficients are non-zero for $i \in
[k-1, k+2n]$ --- two points broader than the delta itself.

We have found that a very sharp delta function, like the two-point
model described in the previous section, leads to instabilities for
orbits with varying $r$ or $\theta$ (e.g., for eccentric
orbits). Using a smoother $n$-point representation suppresses these
instabilities; this will be discussed in greater detail in
\cite{paper2}.  Since $r$ and $\theta$ are constant for circular,
equatorial orbits, these instabilities do not arise for the cases
examined in detail here. Thus for the case at hand, our numerical
errors originating from the finite representation of the delta are
smallest when $n = 1$. This is demonstrated in detail in Sec.\
\ref{sec:results}.

\subsection{Higher order interpolation for smoothness}
\label{sec:cubic}

Finally, we present a representation of the delta which uses a higher
order interpolation scheme.   This again spreads the ``stencil'' of
the delta function over a wider patch of the grid, but improves our
ability to reproduce the integral formulation of the delta identities.

Using cubic interpolation (which requires a total of four points), we
find the rule
\begin{widetext}
\begin{eqnarray}
  h\sum_i f(x_i) \delta_i & = & f(\alpha) \\
	\Rightarrow		 h\sum_i f(x_i) \delta_i		& = &   -\frac{(\alpha-x_{k+1})(\alpha-x_{k+2})(\alpha-x_{k+3})}{6h^3} f(x_k)\nonumber \\
					&  &+\frac{(\alpha-x_k)(\alpha-x_{k+2})(\alpha-x_{k+3})}{2h^3} f(x_{k+1})\nonumber \\
					&   &-\frac{(\alpha-x_k)(\alpha-x_{k+1})(\alpha-x_{k+3})}{2h^3} f(x_{k+2}) \nonumber \\
					&   & +  \frac{(\alpha-x_k)(\alpha-x_{k+1})(\alpha-x_{k+2})}{6h^3} f(x_{k+3})\;.
\end{eqnarray}
The location $\alpha$ lies between grid points $x_{k+1}$ and
$x_{k+2}$.  From this expression, we read off the coefficients
\begin{eqnarray}
\label{cubicdelta}
\-\delta_i & =  &  -\frac{(\alpha-x_{k+1})(\alpha-x_{k+2})(\alpha-x_{k+3})}{6h^4} \mbox{ at } x_k\;,\\
 		 & 	= &	+\frac{(\alpha-x_k)(\alpha-x_{k+2})(\alpha-x_{k+3})}{2h^4}  \mbox{ at } x_{k+1}\;, \\
		 &  = &  -\frac{(\alpha-x_k)(\alpha-x_{k+1})(\alpha-x_{k+3})}{2h^4}  \mbox{ at } x_{k+2} \;,\\
 		 & = &	+  \frac{(\alpha-x_k)(\alpha-x_{k+1})(\alpha-x_{k+2})}{6h^4} \mbox{ at } x_{k+3} \;.
\end{eqnarray} 
A similar analysis for the first derivatives yields
\begin{eqnarray}
\label{cubicdelta_fd}
\delta^\prime_i & = & \frac{(x_{k+1}-\alpha ) (\alpha -x_{k+2}) (\alpha -x_{k+3})}{12 h^5} \mbox{ at } x_{k-1} \;,\\
		 & = & \frac{(\alpha -x_{k}) (\alpha -x_{k+2}) (\alpha -x_{k+3})}{4 h^5} \mbox{ at } x_{k} \;,\\
		 & = & -\frac{(\alpha -x_{k+1}) (2 \alpha -3 x_{k}+x_{k+2}) (\alpha -x_{k+3})}{12 h^5}\mbox{ at } x_{k+1} \;,\\
		 & = & -\frac{(\alpha -x_{k}) (\alpha -x_{k+2}) (2 \alpha +x_{k+1}-3 x_{k+3})}{12 h^5}  \mbox{ at } x_{k+2} \;,\\
		 & = & \frac{(\alpha -x_{k}) (\alpha -x_{k+1}) (\alpha -x_{k+3})}{4 h^5}\mbox{ at } x_{k+3} \;,\\
		 & = & -\frac{(\alpha -x_{k}) (\alpha -x_{k+1}) (\alpha -x_{k+2})}{12 h^5}\mbox{ at } x_{k+4} \;.
\end{eqnarray}
 
Note, we have used the second order finite difference formula, Eq.
(\ref{fd_deriv1}), to derive this result.   In principle, higher order
formulas for the derivative could have been used.   We kept to the
second order formula in order to keep the derivative stencil narrow,
and also for consistency with our time-stepping algorithm.

Finally, for the second derivatives we find
\begin{eqnarray}
\label{cubicdelta_sd}
\delta^{\prime\prime}_i & = & \frac{(x_{k+1}-\alpha ) (\alpha -x_{k+2}) (\alpha -x_{k+3})}{6 h^6} \mbox{ at } x_{k-1}\;,\\
					   & = & \frac{(5 \alpha -3 x_{k}-2 x_{k+1}) (\alpha -x_{k+2}) (\alpha -x_{k+3})}{6 h^6} \mbox{ at } x_{k} \;,\\
					   & = & -\frac{\left(10 \alpha ^2-(9 x_{k}+4 x_{k+1}+7 x_{k+2}) \alpha + x_{k+1} x_{k+2}+3 x_{k} (x_{k+1}+2 x_{k+2})\right) (\alpha
   -x_{k+3})}{6 h^6} \mbox{ at } x_{k+1}\;, \\
   						& = & \frac{(\alpha -x_{k}) \left(10 \alpha ^2-(7 x_{k+1}+4 x_{k+2}+9 x_{k+3}) \alpha +3 x_{k+2} x_{k+3}+ x_{k+1} (x_{k+2}+6
   x_{k+3})\right)}{6 h^6} \mbox{ at } x_{k+2} \;,\\
   						& = & -\frac{(\alpha -x_{k}) (\alpha -x_{k+1}) (5 \alpha -2 x_{k+2}-3 x_{k+3})}{6 h^6} \mbox{ at } x_{k+3} \;,\\
   						& = & \frac{(\alpha -x_{k}) (\alpha -x_{k+1}) (\alpha -x_{k+2})}{6 h^6}\mbox{ at } x_{k+4} \;.
\end{eqnarray}
\end{widetext} 
As discussed in more detail in the following section, our analysis
suggests that this cubic interpolation method works best.
 
We emphasize at this point that, although we are motivated by
Teukolsky equation applications, our discussion here was not
specialized to the Teukolsky equation in any way.  The delta models
sketched here can be used in any finite-difference numerical
algorithm.  We also note that one does not need to stop at cubic-order
interpolation; the basic idea of that scheme could easily be extended
to higher order if the application warranted it.  As the order is
increased, the ``stencil'' of the delta is likewise increased, pushing
us away from the intuitive notion of a structureless impulse.  This
leads us to believe that there may be a certain interpolation order
beyond which the model ceases to work well.

\subsection{Convergence with the discrete delta function}
\label{sec:delta_convergence}

The non-smooth nature of the discrete delta function makes
understanding the convergence properties of a code based on this
function somewhat subtle.  Here we briefly summarize some key issues
related to convergence with the discrete delta.  This summary is based
on detailed discussion of discretization errors given in Ref.\
{\cite{te04}}.  The punchline of this discussion is that the discrete
delta function is typically at least second-order convergent, and thus
we expect our code to likewise be second-order convergent.

Let $\delta_i$ be the discretized version of $\delta(x-\alpha)$
defined on a discrete grid ${x_i}$, let $h = x_{i+1} -x_{i}$ be the
grid spacing, and let $\delta_i$ be non-zero at $x_k, x_{k+1}, \ldots,
x_{k+2n-1}$.  The continuous variable $\alpha$ lies between
$x_{k+n-1}$ and $x_{k+n}$.  This allows us to define a parameter
$\eta$ such that
\begin{equation}
\alpha = x_{k+n-1} + \eta h . 
\end{equation} 
This quantity is a measure of how close $\alpha$ is to a grid point;
clearly, $0 \le \eta \le 1$.

We now define the moments of the discrete delta by
\begin{eqnarray}
M_r(\delta,\alpha,h) = h \sum_{i =k}^{k+2n-1} \delta_i (x_i -
\alpha)^r\;,
\end{eqnarray}
where $r$ is an integer.  In the continuum limit, this definition
becomes
\begin{eqnarray}
M_r &\to& \int \delta(x - \alpha)\left(x - \alpha\right)^r\,dx\;,
\nonumber\\
&=& 1\qquad r = 0
\nonumber\\
&=& 0\qquad r > 0\;.
\end{eqnarray}
A discrete representation will clearly have the correct zeroth moment;
however, it will only have $M_{r > 0} = 0$ up to some maximum $r$.
Reference {\cite{te04}} proves that, if $q$ is the lowest integer such
that $M_q \ne 0$, then
\begin{equation}
\left|f(\alpha) - h\sum \delta_i f(x_i)\right| \le C h^q\;,
\end{equation}
where $C$ is approximately\footnote{In our application, $C$ varies
slightly depending on how close the delta peak is to a grid point.} a
constant.  This delta representation is then $q$th-order convergent.

For the multiple point discrete delta discussed in Sec.\
{\ref{sec:npoint}}, we find $M_0 = 1$, $M_1 = 0$, $M_2 \ne 0$.  When
we use this discrete delta, we therefore expect our code to be
second-order convergent.  We demonstrate this behavior in Sec.\
{\ref{sec:code_convergence}}.  For the cubic delta function, we find
$M_0 = 1$, $M_{1,2,3} = 0$, $M_4 \ne 0$.  In this case, errors due to
the delta representation are expected to be fourth order.  However,
since our stepping algorithm is itself second-order, we expect
second-order convergence overall.

\section{Numerical Implementation and Evaluation of the Discrete
Delta Function}
\label{sec:results}

We now implement the Teukolsky equation's source term using the
techniques discussed in Sec.\ {\ref{sec:delta}} for the simple case of
a point particle in a circular, equatorial orbit around a massive
black hole.  Our goal is to compare the different forms of the
discrete delta discussed in the previous section and to evaluate which
is likely to work best for practical modeling of radiation from
astrophysical systems.   We also compare our discrete delta model to
the Gaussian approximation that has been used in previous work,
illustrating the power of this new model.

We obviously require some ``standard'' to compare our results against.
Frequency-domain codes provide extremely accurate results for
circular, equatorial orbits, largely since their emitted radiation is
concentrated in a small number of multipoles; as such, they make an
excellent standard against which to compare our results.   We use the
code described in~\cite{h2000} as our standard.

\begin{table*}[FDm2]\caption{\label{FDm2}
Energy flux extracted at $R \equiv r_{\rm extract} = 250M$ for
circular, equatorial orbits for the $m=|2|$ mode of a particle with
mass $\mu/M=1$. $a/M$ is the BH spin, $r_0/M$ is the orbital
radius. The labels ``$\delta$'' and ``G'' refer to the results from
$\delta$-code and G-code respectively. Values listed under ``FD'' are
the corresponding fluxes from the frequency-domain code used in
\cite{h2000}.  }
\begin{ruledtabular}
\begin{tabular}{|c|c|c|c|c|c|c|}
    $a/M$    &   $r_0/M$ &  $\dot{E}_{\delta}$ &  $\dot{E}_{FD}$  & $(\dot{E}_{\delta}-\dot{E}_{FD})/\dot{E}_{FD}$ &  $\dot{E}_{G}$ & $(\dot{E}_{G}-\dot{E}_{FD})/\dot{E}_{FD}$\\\hline
  $0$ 	& 	$6$	& $7.385\times 10^{-4}$ & $7.368\times 10^{-4}$ & $0.0023$	& $7.246\times 10^{-4}$ & $-0.017$\\
  $0$     &   $8$   & $1.650\times 10^{-4}$   & $1.651\times 10^{-4}$  & $-0.0055$ & $1.623\times 10^{-4}$ & $-0.016$\\
   $0$     &   $10$   & $5.344\times 10^{-5}$  & $5.374\times 10^{-5}$  & $-0.0004$ &	$5.290\times 10^{-5}$ & $-0.016$\\\hline
   $0.5$     &   $6$   & $5.551\times 10^{-4}$  & $5.539\times 10^{-4}$  & $0.0022$ & $5.437\times 10^{-4}$ & $-0.018$\\
  $0.5$     &   $8$   & $1.399\times 10^{-4}$   &  $1.401\times 10^{-4}$  & $-0.0015$  & $1.375\times 10^{-4}$	& $-0.019$\\
   $0.5$     &   $10$   & $4.781\times 10^{-5}$  & $4.812\times 10^{-5}$  & $-0.0065$  & $4.691\times 10^{-5}$ & $-0.025$\\\hline 
  
    $ 0.9$     &   $4$   & $2.654\times 10^{-3}$   & $2.662\times 10^{-3}$    & $-0.0030$ & $2.611\times 10^{-3}$	& $-0.019$\\
     $0.9$     &   $6$   & $4.614\times 10^{-4}$   &  $4.621\times 10^{-4}$   & $-0.0016$ & $4.531\times 10^{-4}$	& $-0.020$\\ 
   $0.9$     &   $8$   & $1.249\times 10^{-4}$   & $1.254\times 10^{-4}$   &$-0.0039$  & $1.230\times 10^{-4}$ & $-0.019$\\
   $0.9$     &   $10$   & $4.419\times 10^{-5}$  & $4.456\times 10^{-5}$  & $-0.0084$ & $4.339\times 10^{-5}$ & $-0.026$	\\\hline 
      $0.99$     &   $4$   & $2.469\times 10^{-3}$   & $2.484\times 10^{-3}$   & $-0.0059$ &	$2.434\times 10^{-3}$ & $-0.020$\\
     $0.99$     &   $6$   &  $4.450\times 10^{-4}$  & $4.461\times 10^{-4}$   & $-0.0024$	& $4.372\times 10^{-4}$	& $-0.020$\\ 
   $0.99$     &   $8$   &   $1.221\times 10^{-4}$ &  $1.226\times 10^{-4}$  &  $-0.0041$ &	$1.201\times 10^{-4}$ &$-0.020$\\
  $ 0.99$     &   $10$   &  $4.346\times 10^{-5}$ & $4.386\times 10^{-5}$  & $-0.0090$ & $4.270\times 10^{-5}$  &  $-0.026$ \\   
\end{tabular}
\end{ruledtabular}
\end{table*}

\begin{table*}[FDm3]\caption{\label{FDm3}
Energy flux extracted at $250M$ for circular, equatorial orbits for
the $m = |3|$ mode of a particle with mass $\mu/M = 1$.  All symbols
and notation are as in Table \ref{FDm2}.  }
\begin{ruledtabular}
\begin{tabular}{|c|c|c|c|c|c|c|}

 $a/M$    &   $r_0/M$ &  $\dot{E}_{\delta}$ &  $\dot{E}_{FD}$ & $(\dot{E}_{\delta}-\dot{E}_{FD})/\dot{E}_{FD}$ &  $\dot{E}_{G}$ & $(\dot{E}_{G}-\dot{E}_{FD})/\dot{E}_{FD}$\\\hline
  $0$		& $6$ &	$1.465\times 10^{-4}$	&	$1.460\times 10^{-4}$	&	$0.0035$	& $1.431\times 10^{-4}$	& $-0.020$\\
  $0 $	& 	$8$	& $2.445\times 10^{-5}$	& $2.449 \times 10^{-5}$	& $-0.0017$ & $2.399\times 10^{-5}$ & $-0.020$\\
  $0$     &   $10$ & $6.383\times 10^{-6}$ & $6.435\times 10^{-6}$ & $-0.0080$ & $6.291\times 10^{-6}$ & $-0.022$\\\hline
  
   $0.5$     &   $6$   & $1.015\times 10^{-4}$    & $1.014\times 10^{-4}$    & $0.0011$ & $0.992\times 10^{-4}$	&$-0.021$ \\
  $0.5$     &   $8$   & $1.993\times 10^{-5}$   & $1.980\times 10^{-5}$   & $0.0066$  & $1.935\times 10^{-5}$	&$-0.023$\\
   $0.5$     &   $10$   & $5.521\times 10^{-6}$  & $5.572\times 10^{-6}$  & $-0.0090$   & $5.410\times 10^{-6}$ & $-0.029$ \\\hline 
  
     $0.9$     &   $4$   &$6.485\times 10^{-4}$    & $6.467\times 10^{-4}$   &$0.0028$  & $6.336\times 10^{-4}$ & $-0.020$\\
     $0.9$     &   $6$   &$8.031\times 10^{-5}$    &$8.043\times 10^{-5}$    &$-0.0015$ &	$7.865\times 10^{-5}$	& $-0.022$ \\ 
   $0.9$     &   $8$   &  $1.710\times 10^{-5}$  & $1.717\times 10^{-5}$   &$-0.0043$  & $1.677\times 10^{-5}$	& $-0.023$\\
   $0.9$     &  $ 10$   & $4.992\times 10^{-6}$  & $5.044\times 10^{-6}$  & $-0.0103 $&	$4.893\times 10^{-6}$ & $-0.030$ \\\hline 
   
      $0.99$     &   $4$   &  $5.932\times 10^{-4}$  & $5.924\times 10^{-4}$   & $0.0014$ & $5.805\times 10^{-4}$	& $-0.021$\\
     $0.99$     &   $6$   &  $7.688\times 10^{-5}$   & $7.688\times 10^{-5}$   &$-4.8 \times 10^{-5}$ & $7.528\times 10^{-5}$	& $-0.022$ \\ 
   $0.99$     &   $8$   &  $1.6542\times 10^{-5}$  &  $1.669\times 10^{-5}$   & $-0.0086$ & $1.628\times 10^{-5}$	& $-0.025$ \\
   $0.99$     &   $10$   &  $4.879\times 10^{-6}$ &  $4.942\times 10^{-6}$ & $-0.0127$ & $4.792\times 10^{-6}$ & $-0.030$  \\  
\end{tabular}
\end{ruledtabular}
\end{table*}

Tables \ref{FDm2} and \ref{FDm3} shows energy fluxes obtained from our
code for the most dominant azimuthal modes, $|m|=2$ and $3$
respectively. We compare these figures with those obtained from the
code used in~\cite{h2000}\footnote{Symmetry in the azimuthal direction
results in equal fluxes for $+m$ and $-m$ modes. Thus, $|m|$ refers to
the sum of the fluxes from the $+m$ and $-m$ modes which is equal to
twice the flux from either the $+m$ or the $-m$ mode.}.

There are two major reasons that the fluxes we compute depart from
those computed by frequency-domain codes.  First, the time-domain code
must extract fluxes at some finite radius.  The FD approach produces,
by construction, the waveforms and fluxes that would be measured
infinitely far from the generating binary; this simply cannot be done
on a finite radial grid.  A detailed discussion of the impact of
finite extraction radius is given in Sec.\ \ref{sec:rad_extract}.  In
brief, we find by varying the extraction radius that fluxes can be fit
to a very simple power law.  This power law then allows us to infer
the flux that would be measured by distant observers.  The second
source of error is simply numerical --- finite difference errors plus
the approximate nature of our discrete delta.  Roughly speaking,
accounting for finite extraction radius reduces our errors by about a
factor of 2 -- 5; the residual error is thus most likely simply
numerical error.  This is described in much greater detail in Sec.\
\ref{sec:rad_extract}.

\begin{figure}[htb]
\begin{center}
\includegraphics[height = 66mm]{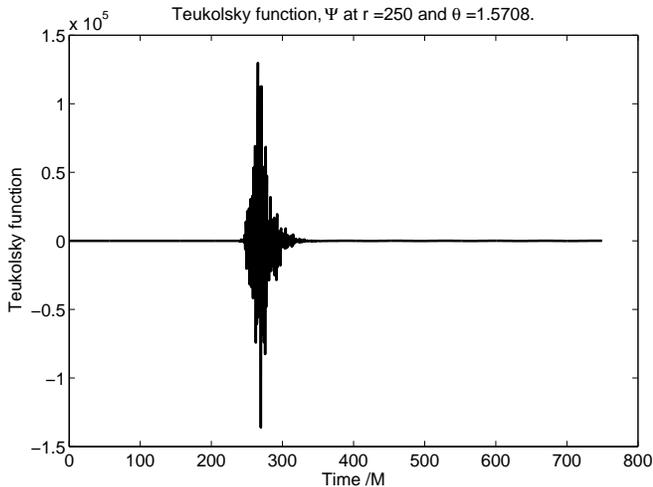}
\caption{The real part of the $m = 2$ mode of the Teukolsky function
$\Psi$ as a function of time for a point particle of mass $\mu/M =
0.01$ in a circular orbit of radius $r_0/M = 7.9456$.   These data
were extracted in the equatorial plane ($\theta = \pi/2$) at radius $R
= 250M$.   At this location the Teukolsky function is zero by
construction until $t \simeq 250M$, at which point a spurious burst
reaches the extraction radius.   This burst is due to our unphysical
initial conditions; it quickly propagates off the grid, leaving a
reasonable physical solution for all time afterwards.\label{burst}}
\end{center}
\end{figure}

\begin{figure}[htb]
\begin{center}
\includegraphics[height = 70mm ]{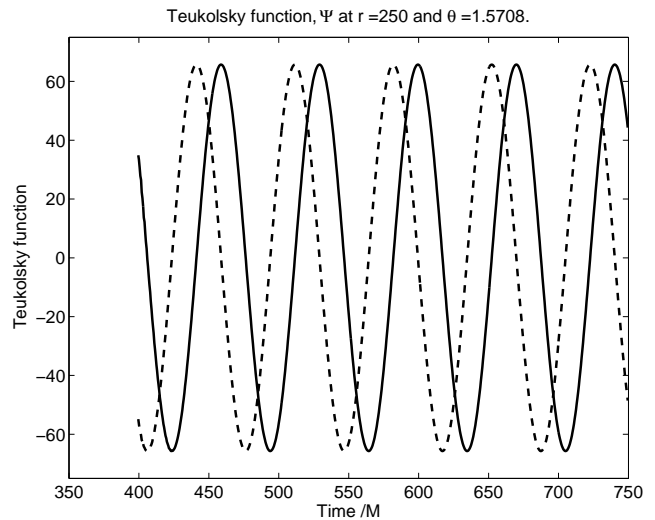}
\caption{ The same as Fig.\ \ref{burst}, but zooming in on the data
for $t \gtrsim 350 M$.   Solid and dashed lines are the real and
imaginary parts of $\Psi$ respectively.   The Teukolsky function
oscillates with a period of about $70M$; since the source has a period
of about $140M$, this is exactly what we expect for the $m = 2$
mode.   We measure the total flux of energy carried by this mode to be
$\dot E/\mu^2 = 1.708 \times 10^{-4}$, in agreement with previous
results (see, e.g., Ref.\ \cite{sl06}).
\label{psiVst}}
\end{center}
\end{figure}

Figures \ref{burst} and \ref{psiVst} illustrate a typical example of
the structure for the Teukolsky function $\Psi$ that we find.   We
show the $m = 2$ mode of an orbit with radius $r_0 = 7.9456M$ around a
Schwarzschild black hole; this orbit was selected in order to compare
with results published in~\cite{sl06}.   Note that the orbital period
at this radius is $T = 2\pi\sqrt{r_0^3/M} \simeq 140M$. The data is
read out at radius $R \equiv r_{\rm extract} = 250M$, $\theta =
\pi/2$; our numerical grid runs from $-100M \le r^* \le 500M$, with a
resolution $\delta r^* = 0.0625M$, and from $0 \le \theta\le \pi$ with
$\delta\theta = \pi/40$.

Figure \ref{burst} shows the real part of $\Psi$ over a broad span of
time, from the beginning of our simulation to $t \simeq 800M$.   At $t
\sim 250M$, a very high amplitude, unphysical burst of radiation
reaches the extraction radius.   This spurious burst is due to our
initial conditions: We initially set $\Psi = 0$ and $\partial_t\Psi =
0$, which is not consistent with our source function.   The time at
which this burst reaches the extraction radius is perfectly consistent
with radiation propagating at the speed of light ($c = 1$ in our
units) across our numerical grid.   The burst quickly propagates off
the grid, and the solution for $\Psi$ settles down to a simple
oscillation. This is shown in Fig.\ \ref{psiVst}, which zooms in on
the behavior of $\Psi$ for $t \gtrsim 350 M$.   Notice that $\Psi_R$
has an oscillation period of about $70M$, precisely what we expect for
the $m = 2$ mode of a source whose orbital frequency is $140M$.   The
energy flux we find from this mode is $\dot E/\mu^2 = 1.708 \times
10^{-4}$, in excellent agreement with results published in
Ref.~\cite{sl06} (compare Table II of~\cite{sl06}, noting that our
results require summing over all $l$ for fixed $m$).

\begin{figure}[htb]
\begin{center}
\includegraphics[height = 90mm]{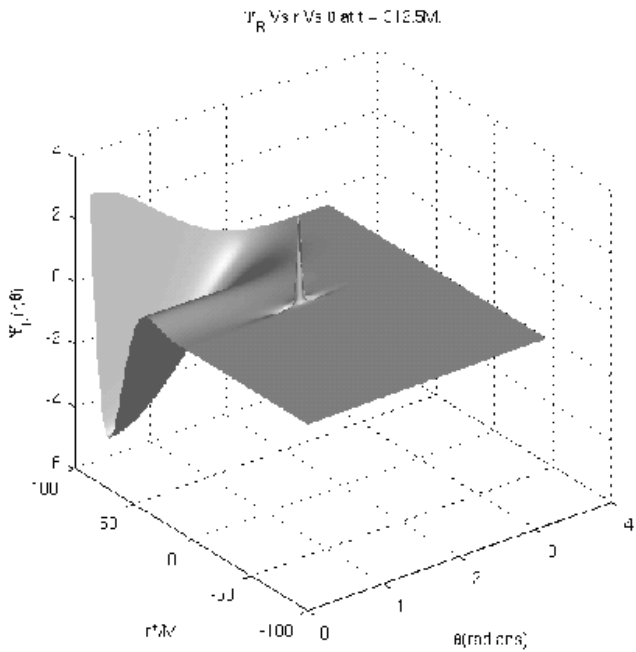}
\caption{The same as Fig.\ 2, but now showing the data for a given
moment in time ($t = 312.5M$) for a wide range of $r^*$ and $\theta$.
Along with the outward propagating radiation packet visible at large
radius, the nearly singular delta function source is clearly visible
at the particle's position.
 \label{psiVsrVsthe}}
\end{center}
\end{figure}

Figure \ref{psiVsrVsthe} illustrates the spatial behavior of the real
part of $\Psi$ at a particular moment in time ($t = 312M$).   This
plot illustrates the behavior of ${\rm Re}\ \Psi$ as a function $r^*$
and $\theta$ over a wide span of our grid.   Clearly visible are the
$m = 2$ mode of the radiation propagating to large radius as well as
the nearly singular delta function source itself.

\subsection{Comparison of different discrete delta functions}
\label{sec:compare}

In this section, we compare results from the various models for the
delta function presented in Sec.\ \ref{sec:delta}. The variable point
approximation for the linear delta function, presented in Sec.\
\ref{sec:delta} provides us with a nice handle to study the
convergence of our results. As we increase $n$, the half-width of the
delta function, the singularity spreads out and its sharpness
decreases. Notice that the physical spread of the source term is
$2(n+1)\delta r^*$(due to the spread of the delta derivatives). Thus,
decreasing the resolution has the same effect on the physical width as
decreasing $n$.

In Tables \ref{variable_delta1} and \ref{variable_delta2}, we present
results describing two equatorial circular orbits, one in the
extremely strong field ($r_0/M = 2.32$, $a/M = 0.9)$, and another at
weaker field ($r_0/M = 12$, $a/M = 0$).  We show the variation in flux
with $n$, the half-width of the radial delta function.  The angular
delta function is represented using two points (i.e., $n_{\rm ang} =
1$), the minimum number of non-trivial points.  The resolutions
$(\delta r_*, \delta\theta)$ are held fixed at $(0.0625M, \pi/40)$

The third and fourth columns of Tables \ref{variable_delta1} and
\ref{variable_delta2} compare the flux in energy carried by GWs as
computed using the $\delta$-code to flux computed using our
frequency-domain standard.  The third column gives a ``raw''
comparison --- we extract the time-domain fluxes at radius $R = 250M$
and compare to the frequency-domain result.  In the fourth column, we
extrapolate the time-domain data, $R \to \infty$, using the algorithm
described in Sec.\ \ref{sec:rad_extract}.  The fourth column thus
contains the most relevant data for assessing which delta
representation is ``best''.  We include the third column to
demonstrate that the difference before extrapolating is not terribly
large, but that it is large enough that the gain due to this
extrapolation is significant.  It also illustrates that not performing
the extrapolation can mislead regarding which form of the delta is
most accurate.

Data from Table \ref{variable_delta2} indicate that, among the
$n$-point representations, $n = 1$ gives the best results for
circular, equatorial orbits.  The cubic representation, however, is
even better --- the smoothness of this technique apparently reduces
error even more.  We choose the cubic delta for the remainder of our
analysis because it is both smooth and accurate.

By contrast, the most accurate flux in Table \ref{variable_delta1}
occurs when $n = 7$ (total of 8 points to represent the source),
rather than $n = 1$.  The reason is due to a competition between
errors from finite differencing and errors from our delta
representation.  In particular, we have noticed experimentally that
finite difference errors tend, on average, to spuriously decrease our
measured flux; errors from spreading the delta over the grid tend to
augment the flux.  (We emphasize that this is merely a rule-of-thumb
tendency we have noted; we also emphasize that we do not as of yet
have a good explanation for these effects.)

As we approach the horizon, finite difference errors tend to become
more important.  This can be compensated by increasing the width of
our delta representations.  At $n = 7$, the spread of our source is
just enough to accurately compensate for finite difference errors.  At
larger radius (e.g., the $r_0 = 12 M$ orbit shown in Table
\ref{variable_delta2}), finite difference errors are so small that we
do best using the minimum number of points possible in our model.

\begin{table*}[variable_delta1]
\caption{\label{variable_delta1}Comparison of several implementations
of the discrete delta function.   We show results for the linear hat
delta described in Sec.\ {\ref{sec:npoint}}, as well as the cubic
delta function described in Sec.\ {\ref{sec:cubic}}.   All fluxes are
measured at $R = 250M$ for the $|m| = 2$ mode. For the linear hat
delta, the total number of points in the function is $2(n+1)$.  The
cubic delta uses 6 points in all. These results are for orbits of
radius $r_0 = 2.32M$ about a black hole with $a = 0.9M$.   The total
flux in $|m| = 2$ modes according to our frequency-domain standard is
$\dot E_{\rm FD}/\mu^2 = 2.061 \times 10^{-2}$. }
\begin{tabular}{|c|c|c|c|c|}
\hline
Total points, $2(n+1)$	&	$\dot{E}_{250} $	&	$(\dot{E}_{250}-\dot{E}_{FD})/\dot{E}_{FD}$ & $\dot{E}_{\infty} $ & $(\dot{E}_{\infty}-\dot{E}_{FD})/\dot{E}_{FD} $\\
\hline
$64$	&	$2.889 \times 10^{-2}$ & $4.0 \times 10^{-1}$& $2.890 \times 10^{-2}$ & $4.0 \times 10^{-1}$\\
$32$	&	$2.194 \times 10^{-2}$ & $6.5 \times 10^{-2}$& $2.195 \times 10^{-2}$ & $6.5 \times 10^{-2}$\\
$16$	&	$2.055 \times 10^{-2}$ & $-2.8 \times 10^{-3}$& $2.056\times 10^{-2}$ & $-2.4\times 10^{-3}$\\
$8$	&	$2.027 \times 10^{-2}$ & $-1.6 \times 10^{-2}$& $2.028 \times 10^{-2}$ & $-1.6 \times 10^{-2}$\\
$4$	&	$2.023 \times 10^{-2}$ & $-1.9 \times 10^{-2}$& $2.024 \times 10^{-2}$ & $-1.8 \times 10^{-2}$\\
cubic  &	$2.024 \times 10^{-2}$	& $-1.8 \times 10^{-2}$ & $2.024 \times 10^{-2}$ & $-1.8\times 10^{-2}$ \\
\hline
\end{tabular}
\end{table*}

\begin{table*}[variable_delta2]
\caption{\label{variable_delta2}Same as Table \ref{variable_delta1},
but now for an orbit with $r_0 = 12M$ about a black hole with $a = 0$.
The frequency-domain flux for $|m| = 2$ modes in this case is $\dot
E_{\rm FD}/\mu^2 = 2.172 \times 10^{-5}$.}
\begin{tabular}{|c|c|c|c|c|}
\hline
Total points, $2(n+1)$	&	$\dot{E}_{250} $	&	$(\dot{E}_{250}-\dot{E}_{FD})/\dot{E}_{FD}$ & $\dot{E}_{\infty} $ & $(\dot{E}_{\infty}-\dot{E}_{FD})/\dot{E}_{FD} $\\
\hline
$64$	&	$2.342 \times 10^{-5}$ & $7.8\times 10^{-2}$ & $2.376\times 10^{-5}$ & $9.4\times 10^{-1}$\\
$32$	&	$2.191\times 10^{-5}$ & $8.7\times 10^{-3}$ & $2.224\times 10^{-5}$ & $2.4\times 10^{-1}$\\
$16$	&	$2.156\times 10^{-5}$ & $-7.5\times 10^{-3}$ & $2.187\times 10^{-5}$ & $7.1\times 10^{-3}$\\
$8$	&	$2.148\times 10^{-5}$ & $-1.1\times 10^{-2}$ & $2.179\times 10^{-5}$ & $3.3\times 10^{-3}$\\
$4$	&	$2.146\times 10^{-5}$ & $-1.2\times 10^{-2}$& $2.177\times 10^{-5}$ & $2.5\times 10^{-3}$\\
cubic  & $2.145\times 10^{-5}$ & $-1.2\times 10^{-2}$ & $2.177\times 10^{-5}$ & $2.3\times 10^{-3}$	\\
\hline
\end{tabular}
\end{table*}

\subsection{Convergence of our code}
\label{sec:code_convergence}

As discussed in Sec.\ {\ref{sec:delta_convergence}}, we generally
expect a code built using the discrete delta on grid with spacing $h$
to exhibit $O(h^2)$ convergence.  In particular, we expect the Weyl
scalar $\psi_4$ to show second-order convergence.  We check this
expectation by examining the flux of energy carried by gravitational
waves.  Since we expect $\psi_4 = \psi_4^{\rm true} + O(h^2)$, we
likewise expect $\dot E$ to exhibit second-order convergence:
\begin{equation}
\dot E \sim |\psi_4|^2 \sim |\psi_4^{\rm true}|^2 + O(h^2)\;.
\end{equation}

\begin{figure}[htb]
\begin{center}
\includegraphics[height = 90mm]{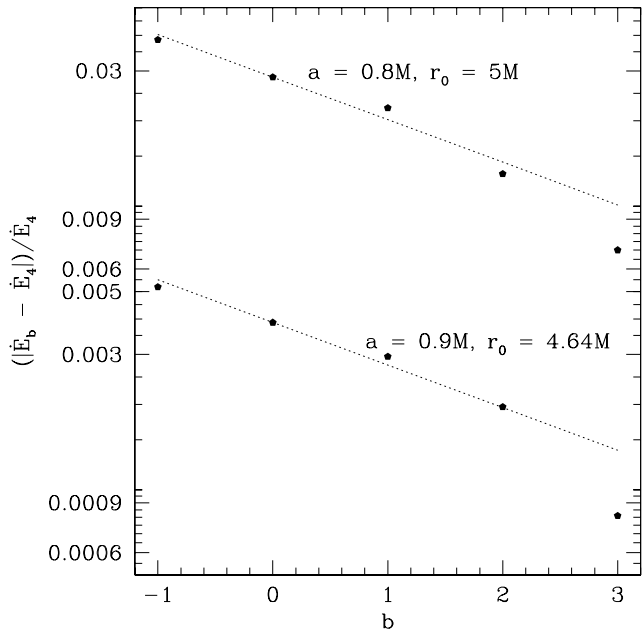}
\caption{An ilustration of our code's convergence behavior.  We show
the fractional deviation in energy flux in the $|m| = 2$ mode,
measured at $R = 250M$ as a function of grid spacing.  The grid is
controlled by the integer $b$ using $\delta r^* = 0.0625\times
2^{-b/4}$, $\delta\theta = \pi/30\times 2^{-b/4}$, with $b \in
[-1,\ldots,4]$.  The upper data set is for fluxes measured from an
orbit with $a = 0.8M$, $r_0 = 5M$; the lower set is for $a = 0.9M$,
$r_0 = 4.64M$.  For each data set, the dotted line represents what we
would expect for perfect second-order convergence (fit arbitrarily to
the data for $b = 0$); the large dots represent our actual convergence
data.
\label{fig:conv}}
\end{center}
\end{figure}

To demonstrate this convergence, we show the energy flux measured at
$R = 250M$ for two different strong-field\footnote{It's worth noting
that we found it to be rather difficult to demonstrate convergence
using weak-field orbits.  For such orbits, the differences in our
computed fluxes were quite small as we varied our grid density.  We
need strong-field orbits in order for the errors to be large enough
that the convergence trend is apparent.}  orbits: $r_0 = 5M$, $a =
0.8M$ and $r_0 = 4.64M$, $a = 0.9M$.  The radial and angular grids are
set to
\begin{eqnarray}
\delta r^* = 0.0625 \times 2^{-b/4}\;,
\\
\delta\theta = \pi/30 \times 2^{-b/4}\;.
\end{eqnarray}
Actually, $\delta\theta$ was modified slightly from this to insure
that $\theta = \pi/2$ lies exactly on a grid point.  This reduced
variations about the main $h^2$ trend owing to the slight dependence
of the proportionality ``constant'' on the delta's peak (cf.\
discussion in Sec.\ {\ref{sec:delta_convergence}}).  Figure
{\ref{fig:conv}} shows the results our runs for $b \in [-1,\ldots,4]$.
Convergence is shown by examining the fractional error with respect to
our densest grid,
\begin{equation}
\mbox{error} \equiv \frac{|\dot E_b - \dot E_4|}{\dot E_4}\;,
\end{equation}
where $\dot E_b$ is the flux inferred at grid parameter $b$.  We
normalize to $b = 4$ since it is the densest grid available to us.
Modulo some slight oscillations, the overall trend of our data is in
very good agreement with second-order convergence.

\subsection{Comparison of discrete and Gaussian approximations for
the numerical delta}
\label{sec:compare_disc_gauss}

The work in~\cite{akp03,k04, bk07} approximates the delta-function by
a narrow Gaussian such that it integrates to unity over the numerical
grid. The Gaussian smears out the singularity and thus the source term
is non zero at a large number of points on the numerical grid. In
contrast, the models presented in Sec.\ \ref{sec:delta} use only a few
points to depict the delta-function and thus do not share this
disadvantage. A comparison on the same hardware and software platform
showed that the techniques used here (the $\delta$-code) are about
{\em twelve} times faster than the ones that use a smeared Gaussian
(the G-code). The last two columns in Tables \ref{FDm2} and \ref{FDm3}
show the fluxes from the Gaussian-approximated delta function. Note
that the errors in these fluxes are about $2-3\%$, quite a bit higher
than errors from the $\delta$-code. Both codes were run with identical
parameters and grid resolutions. The accuracy of both codes improves
with higher resolution, and improvement in both is seen by moving the
extraction radius farther out.  However, when these parameters are
fixed, we find that the $\delta$-code is faster and demonstrates
higher accuracy.

\section{Accounting for finite extraction radius}
\label{sec:rad_extract}   

When one discusses the gravitational-wave fluxes which a system
generates, one is normally interested in their asymptotic value
infinitely far away.  It is of course not possible for a finite
coordinate grid to reach all the way into this distant zone, so it is
of great importance to understand how our fluxes vary with respect to
our finite extraction radius $R$.
 
In flat spacetime, the extraction radius is not very important; it
just needs to be sufficiently far away that the field it measures is
purely radiative (i.e, not contaminated by near-field
effects). Conservation laws guarantee that fluxes follow a $1/r^2$ law
in this region, and so the integrated flux is independent of
extraction radius.

Things are not so simple in a curved spacetime --- radiation is
effectively scattered off of spacetime curvature, modifying its
propagation characteristics compared to flat spacetime intuition. This
is responsible for the late time ``tails'' that are seen when a
radiation packet propagates away from a black hole (cf. the late time
behavior seen in Fig.\ \ref{qnm}). These tails can be regarded,
heuristically, as radiation whose propagation to large radius was
delayed by scattering off the spacetime.  It also causes the
integrated flux to depend on and vary with the radius at which it is
measured.
 
We now examine how our fluxes vary with respect to extraction radius.
Tables \ref{ext_rad1} and \ref{ext_rad2} present the fluxes measured
for four representative strong-field orbits.  In each case, we measure
$\dot E$ for the $|m| = 2$ and $|m| = 3$ modes at extraction radii
$R/M = 100$, 200, 300, 400, 500, and 600.  These data are then fit to
the ansatz
\begin{equation}
\dot E = \dot E_\infty\left[1 - q\left(m\Omega_{\rm
orb}R\right)^{-p}\right]\;.
\label{eq:best_fit}
\end{equation}
The parameters $q$, $p$, and $\dot E_\infty$ are determined by the
fit.  Notice that $\dot E_\infty$ represents the flux that (according
to this ansatz) would be measured infinitely far away.  Note that this
form was suggested to us by L.\ M.\ Burko {\cite{lior_private}}, and
replaces a previous version which used $(r_{\rm orb}/R)^p$ rather than
$(m\Omega_{\rm orb}R)^{-p}$.  The two forms can be easily related to
one another; however, the form involving $m\Omega_{\rm orb}$
emphasizes that it is the asymptotic behavior of the mode, rather than
a property of the orbit, that sets $\dot E$.  This form should also be
more readily extendable to non-circular orbits.

\begin{table*}[ext_rad1]
\caption{\label{ext_rad1} Fluxes extracted at a sequence of radii on
the numerical grid. $a/M$ is the BH spin, $r_0/M$ is the orbital
radius and $|m|$ is the azimuthal mode. $\dot{E}_R$ is the flux
measured at radius $RM$. }
\begin{tabular}{c}
\hline
\end{tabular}
\begin{ruledtabular}
\begin{tabular}{|c|c|c|c|c|c|c|c|c|}
 
$|m|$ &$a/M$	& $r_0/M$ &  $\dot{E}_{100}$	& $\dot{E}_{200}$ & $\dot{E}_{300}$ &	$\dot{E}_{400}$ & $\dot{E}_{500}$ & $\dot{E}_{600}$ \\
\hline
$2$	& $0.99$	& $4$ & $2.4567\times 10^{-3}$ & $2.4681\times 10^{-3}$ & $2.4702\times 10^{-3}$ & $2.4709\times 10^{-3}$ & $2.4712\times 10^{-3}$ & $2.4714\times 10^{-3}$  \\
$2$ 	& $0.99$ & $10$	  & $4.1032\times 10^{-5}$ & $4.3209\times 10^{-5}$ & $4.3598\times 10^{-5}$ & $4.3767\times 10^{-5}$ & $4.3828\times 10^{-5}$ & $4.3861\times 10^{-5}$ \\
$2$ 	& $0.90$ & $10$	  & $4.1729\times 10^{-5}$ & $4.3930\times 10^{-5}$ & $4.4322\times 10^{-5}$ & $4.4456\times 10^{-5}$ & $4.4517\times 10^{-5}$ & $4.4550\times 10^{-5}$ \\
$2$	& $0.00$ & $12$	  & $1.9584\times 10^{-5}$ & $2.1256\times 10^{-5}$ & $2.1554\times 10^{-5}$ & $2.1654\times 10^{-5}$ & $2.1699\times 10^{-5}$ & $2.1723\times 10^{-5}$ \\
\hline
$3$	& $0.99$	& $4$ 	& $5.8962\times 10^{-4}$ & $5.9278\times 10^{-4}$ & $5.9334\times 10^{-4}$ & $5.9353\times 10^{-4}$ & $5.9361\times 10^{-4}$ & $5.9364\times 10^{-4}$ \\
$3$	& $0.99$ 	& $10$  & $4.5778\times 10^{-6}$ & $4.8558\times 10^{-6}$ & $4.9051\times 10^{-6}$ & $4.9220\times 10^{-6}$ & $4.9297\times 10^{-6}$ & $4.9339\times 10^{-6}$ \\
$3$ 	& $0.90$	& $10$	& $4.6791\times 10^{-6}$ & $4.9588\times 10^{-6}$ & $5.0085\times 10^{-6}$ & $5.0255\times 10^{-6}$ & $ 5.0333\times 10^{-6}$ & $  5.0375\times 10^{-6}$ \\ 
$3$	& $ 0.00$ 	& $12$	& $1.9528\times 10^{-6}$ & $2.1326\times 10^{-6}$ & $2.1650\times 10^{-6}$ & $2.1761\times 10^{-6}$ &  $2.1812\times 10^{-6}$ &  $2.1839\times 10^{-6}$ \\

\end{tabular}
\end{ruledtabular}
\end{table*}
  
\begin{table*}[ext_rad2]
\caption{\label{ext_rad2} Best fit parameters, $\dot{E}_\infty$, $p$,
$q$ [appearing in Eq.\ (\ref{eq:best_fit})] for data presented in
Table \ref{ext_rad1}. }
\begin{tabular}{c}
\hline
\end{tabular}
\begin{ruledtabular}
\begin{tabular}{|c|c|c|c|c|c|c|c|}
$|m|$ &$a/M$	& $r_0/M$  & $\dot{E}_\infty$ & $p$ & $q$ & $\dot{E}_{FD}$ & $(\dot{E}_\infty-\dot{E}_{FD})/\dot{E}_{FD}$ \\
\hline
$2$	& $0.99$	& $4$  & $2.4718 \times 10^{-3}$ & $2.04 $ & $3.40 $ & $2.4836\times 10^{-3}$ & $-0.0048 $ \\
$2$ & $0.99$ & $10$	 & $4.3953 \times 10^{-5}$ & $1.96 $ & $2.31$ & $4.3948\times 10^{-5}$ & $0.0001 $ \\
$2$ 	& $0.90$ & $10$	 & $4.462 \times 10^{-5}$ & $2.05 $ & $2.70 $ & $4.4560\times 10^{-5}$ & $0.0014 $ \\
$2$	& $0.00$ & $12$	   & $2.1779 \times 10^{-5}$   & $2.07 $ & $2.59 $ & $2.1722\times 10^{-5}$ & $0.0026 $\\
\hline
$3$	& $0.99$	& $4$  & $5.9375 \times 10^{-4}$ & $2.09 $ & $10.74 $ & $5.9239\times 10^{-4}$ & $0.0023 $ \\
$3$	& $0.99$ 	& $10$ & $4.9428 \times 10^{-6}$ & $2.06 $ & $7.27 $ & $4.9417\times 10^{-6}$ & $ 0.0002$ \\
$3$ 	& $0.90$	& $10$	& $ 5.0466 \times 10^{-6}$ & $2.06 $ & $7.16 $ & $ 5.0440\times 10^{-6}$ & $0.0005$ \\ 
$3$	& $ 0.00$ 	& $12$ & $ 2.1900\times 10^{-6}$ & $2.05 $ & $6.20 $ & $2.1890\times 10^{-6}$ & $0.0004 $\\

\end{tabular}
\end{ruledtabular}
\end{table*}

\begin{figure}[htb]
\begin{center}
\includegraphics[height = 75mm]{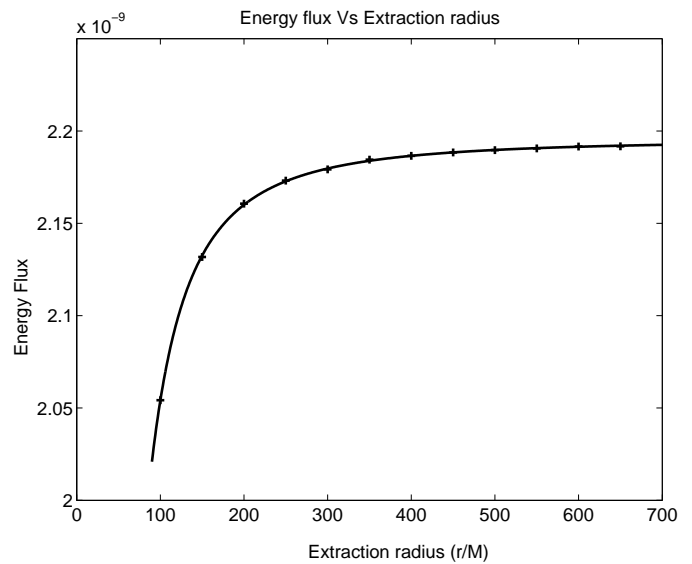}
\caption{A power law fit to our numerically extracted energy fluxes
for the case $r_0/M = 10$, $a/M = 0.99$, $m = +2$.  Numerical data is
indicated by the dots; the curve is the best fit we obtain for the
ansatz given by Eq.\ (\ref{eq:best_fit}).  For this case, the best fit
parameters are $q = 7.45$, $p = 2.06$, $\dot E_\infty = 2.197\times
10^{-5}$.  \label{best_fit}}
\end{center}
\end{figure}

Figure 5 shows an example of how well this fit works for one of the
cases given in Table \ref{ext_rad1} ($r_0/M = 10$, $a/M = 0.99$, $m =
2$).  Pragmatically, this ansatz appears to fit the data quite well;
the quality shown in Fig.\ \ref{best_fit} is typical for the data that
we examined.  Interestingly, we find in all cases that the exponent $p
\simeq 2$, independent of black hole spin, orbit radius, or mode
number.  Detailed calculations which we will present elsewhere
{\cite{us_inprep}} shows that this corresponds to the {\it dominant}
correction to the asymptotic behavior of $\psi_4$.  Such behavior was
also demonstrated by Newman and Unti {\cite{nu62}} (although our
results do not currently agree with theirs when $a \ne 0$; we are
investigating this discrepancy).

We now use the fit (\ref{eq:best_fit}) to compare the extrapolated
measured flux $\dot E_\infty$ to frequency-domain results $\dot E_{\rm
FD}$.  This is shown in the last column of Table \ref{ext_rad2}.  In
all cases, the error we find is less than $1\%$, sometimes
substantially less. A similar fit can be performed on the angular
momentum flux, with similar results.  We conclude that the fit
(\ref{eq:best_fit}) accounts for finite extraction radius, providing
an accurate estimate for the fluxes that a particle radiates to
infinity.  Residual errors are thus much more likely to be true
measures of numerical error in our calculation, and not an artifact of
the extraction.

\section{Summary and Future work}
\label{sec:summary}

We have presented a simple, new technique for modeling the Dirac delta
function and its derivatives on a finite difference grid.  This
technique requires that the source be modeled only on a handful of
points on the grid.  Our particular goal in this analysis is to model
a pointlike source function for the time-domain Teukolsky equation,
appropriate to describe the smaller member of an extreme mass ratio
binary.  We emphasize that our models for the discrete delta and its
derivatives are more broadly applicable than just the Teukolsky
equation --- these techniques can be used in any context that requires
modeling a sharp, delta-like function on a finite difference grid.

We test this approach by solving the Teukolsky equation for a test
body in a circular, equatorial orbit of a Kerr black hole.  Comparing
with a well tested time-domain code that treats the orbiting body
using a truncated Gaussian, we find that this new approach is
extremely fast (often by a factor of $\sim10$) and accurate.  Using a
frequency-domain code as a benchmark to compare the flux of energy
carried by gravitational waves, we find that the code which uses the
discrete delta function is typically a factor of $2 - 5$ more accurate
than the Gaussian treatment most commonly used previously.  This
accuracy can be improved still further (at least for fluxes) by using
a simple fit that accounts for the variation of the flux with the
extraction radius.  Combining our new source function with this
fitting law, we find that our code agrees with the frequency-domain
benchmark with errors smaller than $1\%$ for a large fraction of
parameter space, sometimes significantly smaller.

Since the goal of this analysis is to contribute to the modeling of
EMRI gravitational-wave sources, the restriction to circular and
equatorial orbits, though a useful, illustrative test, is not
astrophysically realistic.  Since such binaries form through
scattering processes, they are expected to have substantial
eccentricity {\cite{gair04}}, and the secondary's orbit should have no
special alignment with the spin axis of the large black hole.  Our
next analysis will study how well this new technique handles such
orbits.  This realistic case is substantially more difficult to treat
than circular, equatorial orbits, since the orbiting body (and our
discrete delta model) very rapidly crosses back and forth over
gridpoints in both the radial and latitudinal directions.  As this
paper is being completed, this work is well underway.  Early analysis
points to excellent results --- we get very good results even when we
move our discrete delta model rapidly in a dynamical orbit.  We also
plan to examine wave emission from non-geodesic trajectories (in
preparation for using the code to calculate wave emission from an
inspiral sequence), and to examine plunging orbits (with a plan to
examine waves from the transition between late inspiral to final
plunge {\cite{orithorne}}).
 
The final goal of this work will be to compute adiabatic inspiral
waveforms using a hybrid of frequency-domain and time-domain, as
described in the introduction.  With a robust time-domain code for
computing waves from nearly arbitrary physical worldlines and with a
robust frequency-domain code capable of ``mass producing'' radiation
reaction data for generic Kerr black hole orbits this problem should
boil down to simple a matter of available CPU resources.  Once we are
in this state, we hope to produce waveforms efficiently enough that
they can be used by workers looking at problems in {\it LISA} data
analysis and waveform measurement (e.g., the ``Mock {\it LISA} Data
Challenge'' {\cite{mock1,mock2,mock3,mock4}}).  These waveforms are
likely to be useful for other astrophysical problems, such as
computing radiation recoil from both the slow inspiral and the dynamic
plunge.  Computing this effect in the extreme mass ratio limit may
serve as a precision check on recent work looking at this problem in
full numerical relativity {\cite{kick1,kick2,kick3,kick4,kick5}}.

\acknowledgments

We thank Lior M.\ Burko for very helpful discussions on various aspects of
this work, particularly for suggesting the orbital-frequency
dependence used in Eq.\ (\ref{eq:best_fit}), and other discussions
regarding the the impact of finite extraction radius on $\dot E$.  We
likewise thank Eric Poisson for helpful discussion regarding finite
extraction radius.  Many of the numerical simulations presented here
were performed at the San Diego Supercomputing Center.  GK
acknowledges research support from the University of Massachusetts and
Glaser Trust of New York, as well as supercomputing support from the
TeraGrid (grant number TG-PHY060047T). PAS and SAH are supported by
NASA Grant No.\ NNG05G105G; SAH is in addition supported by NSF Grant
No.\ PHY-0449884, and gratefully acknowledges the support of the MIT
Class of 1956 Career Development fund.


\begin{thebibliography}{99}
\bibitem{fh_inprep} E.\ E.\ Flanagan, and T.\ Hinderer, in
  preparation.  This paper uses a ``two timescale'' analysis to
  systematically examine the validity of the adiabatic approach to
  inspiral in Kerr spacetimes; it finds and corrects an error in the
  conceptual foundation of Appendix A of Ref.\ {\cite{dfh05}}, but
  shows that the numerical impact of this error does not change that
  reference's conclusion.
\end{thebibliography}
\end{document}